\documentclass[11pt,reqno]{amsart}
\usepackage[colorlinks]{hyperref}
\hypersetup{colorlinks=true,linkcolor={magenta},pdfborder=0 0 0}
\usepackage{rotating}
\usepackage{enumerate}
\usepackage[shortlabels]{enumitem}
\usepackage{tabularx}
\usepackage{array}
\usepackage{makecell}
\usepackage{adjustbox}
\usepackage{longtable}
\usepackage{graphicx}
\usepackage{amssymb}
\usepackage{multirow}
\usepackage{amsmath,amsfonts}%
\numberwithin{equation}{section}
\usepackage{amsthm}%
\usepackage{mathrsfs}%
\usepackage[table]{xcolor}
\definecolor{lavender}{rgb}{0.9,0.9,0.98}%
\usepackage{manyfoot}%
\usepackage{booktabs}%
\usepackage{algorithm}
\usepackage{algorithmicx}%
\usepackage{algpseudocode}%
\usepackage{color,soul}
\usepackage{listings}%
\usepackage{csquotes}
\usepackage{fontawesome}
\usepackage{blindtext}
\usepackage{hyperref}
\usepackage{enumitem}
\usepackage{nccmath} 
\usepackage{nicefrac}       %
\usepackage{microtype}      %
\usepackage[utf8]{inputenc}
\usepackage{nicefrac}       %
\usepackage{microtype}  
\usepackage{float}
\usepackage{natbib}
\usepackage{algpseudocode}
\usepackage[table]{xcolor}
\usepackage{lscape}
\usepackage{fdsymbol}
\usepackage{threeparttable}
\usepackage{wasysym}
\usepackage{fontawesome}
\usepackage{algpseudocode}
\usepackage{algorithm}
\usepackage{multirow}

\newcolumntype{C}[1]{>{\centering\arraybackslash}p{#1}}

\usepackage{comment}
\usepackage{fancyhdr}
\begin{document}
\title{A Method for Securely Transmitting Large Video Files Using Chaotic Compression and Encryption}

\author{Shiladitya~Bhattacharjee, Subha~Bhattacharya, Arnab~Chatterjee,  Sulabh~Bansal, and Saurabh Shukla}
	
	\address{Shiladitya Bhattacharjee, {\tt shiladitya.bhattacharjee@ddn.upes.ac.in}, School of Computer Science, UPES, Dehradun, India.}
    \address{Subha Bhattacharya, {\tt subha.bh.juee26@gmail.com},
    Department of Electrical Engineering, Jadavpur University, Kolkata, West Bengal, India.}
	\address{Arnab Chatterjee, {\tt arnab.chatterjee@upes.ac.in}, School of Computer Science, UPES, Dehradun, India.}
	\address{Sulabh Bansal, {\tt sulabh.bansal@jaipur.manipal.edu}, School of Computer Science and Engineering, Manipal University Jaipur (MUJ), Jaipur, Rajasthan, India.}
	\address{Saurabh Shukla, {\tt saurabh.shukla@iiitl.ac.in}, Department of Computer Science, Indian Institute of Information Technology Lucknow (IIITL), Lucknow, Uttar Pradesh, India.}

\begin{abstract}
  Conventional techniques for compression and encryption are frequently laborious and resource-intensive, rendering them inappropriate for real-time applications.  A plethora of research has been presented in the current literature to address these difficulties together; yet, it fails to propose any suitable strategy. Therefore, this study introduces an innovative simultaneous data compression and encryption (SDCE) system specifically designed for large video files.  The methodology amalgamates chaotic map-based encryption with Huffman encoding for lossless compression into a cohesive framework, markedly diminishing computational overhead and processing duration while augmenting data security.  The logistic map is utilized to produce a pseudo-random chaotic sequence for XOR-based encryption, guaranteeing robust security against unwanted access. The research findings demonstrate its efficacy in enhancing data privacy compared to other existing and related strategies, particularly in terms of generating greater entropy and avalanche effects.  It produces superior throughput, compression ratio, peak signal-to-noise ratio (PSNR), and reduced bits per rate (BPC), along with a smaller percentage of data loss, which further supports its ability to provide enhanced data integrity compared to other existing methods.
  
  \vspace{5mm}
  Keywords: Simultaneous data compression and encryption (SDCE), chaotic map, Huffman encoding,  entropy and avalanche effects, throughput, peak signal-to-noise ratio (PSNR), data loss.
\end{abstract}
\maketitle
\thispagestyle{empty} 
\hypertarget{sec1}{}
\section{Introduction}
Next-generation video storage and network systems are rapidly evolving due to the growth of high-definition video content, various mobile video systems, streaming platforms, and sensor networks. Video files account for a large part of the world's data traffic and are growing at an unprecedented rate, reaching terabytes, petabytes, and even exabytes. This explosive growth disrupts the operational capabilities of individual operations and requires the development of effective and efficient solutions \cite{kar2018improved, tabash2019efficient}. In any extensive data generation model \cite{khan2014big, gupta2025big} and \cite{jadhav2025optimal}, if the initial data volume is represented by $V_{\emptyset}$, and the data volume at a specific time is represented by $V(\theta)$, then $V(\theta)$ may be determined using \hyperlink{eqn1}{Equation (1)}.

\hypertarget{eqn1}{\begin{equation} \tag{1}
 V(\theta)=V_{\emptyset} \times (2)^{\theta/\tau }
\end{equation}}

In \hyperlink{eqn1}{Equation (1)}, $\tau$ denotes the duration necessary for the input data to double. According to the literature \cite{khan2014big, gupta2025big}, if $\eta$ denotes the number of features in the input data, the output data $\digamma$ may be computed using the 
\hyperlink{eqn2}{Equation (2)}.

\hypertarget{eqn2}{}
\begin{equation}
\digamma =(\eta \times \zeta)+ \varphi 
\end{equation}
The variable $\zeta$ represents the effects of $\eta$ on $\digamma$, while $\varphi$ denotes any errors that occur during data production or transmission as outlined in \hyperlink{eqn2}{Equation(2)}. During the handling of operations on a wide scale, the foremost impact is compromised multimedia data privacy and integrity, basically attributed to the security problems involved. Multitudinous trials have been accepted to address the forenamed security difficulties, particularly securing against data compression. and security generally, the increased confines of keys, initialisation vectors, and expansive prosecution cycles contribute to enhanced data sequestration \cite{preishuber2018depreciating}. Still, this also results in significant time and space charges. Hence, the application of intricate ways to cover data sequestration during the transfer of all data or apps is rendered inoperable due to the significant increase in the quantum of these data or operations. Consequently, it results in data loss and compromises data integrity \cite{singh2021level}. The manipulation of data size may be effectively regulated by enforcing several effective data compression techniques. However, the literature indicates that the use of data compression may reduce the information, which may be lost forever. Amongst the several data contraction styles, the lossy compression method  permanently removes a significant proportion of spare data. The data quality is reduced since similar data cannot be recovered during relaxation \cite{song2019efficient,huang2025efficient}. The acceptability of the employed lossy compression technique is significantly influenced by the quality of the retrieved video data in any lossy compression. \hyperlink{eqn3}{Equation (3)} and \hyperlink{eqn4}{Equation (4)} can be used to calculate mean squared error (MSE) and peak signal-to-noise ratio (PSNR) in order to quantify lost or distorted video data \cite{matin2021video, bhattacharjee2016security}.

\hypertarget{eqn3}{\begin{equation} \tag{3}
 MSE={\frac{1}{\upsilon}} \times \sum_{i=1}^{\upsilon}({\alpha}_i -{\delta}_i)
\end{equation}}

In \hyperlink{eqn3}{Equation 3}, $({\alpha_i})$ represents the precise amount of the input video data, and $({\delta_i})$ represents the output video data following any lossy compression; $\upsilon$ represents the total number of input frames on which the lossy compression is applied. The PSNR can be computed using the following \hyperlink{eqn4}{Equation(4)} if $\vartheta$ is the greatest value that any visual signal can have.
\hypertarget{eqn4}{\begin{equation} \tag{4}
PSNR= 10 \times \log_{10}\left(\frac{\vartheta^2}{MSE}\right)
\end{equation}}
If a video processing method yields a higher $PSNR$ number, it means that the video output is of greater quality. The lossless compression technique is also more preferable for data management in the case of multimedia files. According to the `Shannon theory', any lossless compression method's efficiency in preserving the quality of any video data is indicated by the computation of its entropy value \cite{bhattacharjee2015lossless, bhattacharjee2024leveraging, long2025improved} . The following \hyperlink{eqn5}{Equation (5)} can be used to calculate the entropy value ($\Delta$) if the probability of any video frame ($\psi$) is $P(\psi)$, and the total number of video frames is  $\kappa$. 
\hypertarget{eqn5}{\begin{equation} \tag{5}
\Delta= \sum_{i=1}^{\kappa} P(\psi_i) \times \log_{2}{P(\psi_i)}
\end{equation}}

\hyperlink{Fig1}{Fig. 1} shows a broad impression of the encoding of videos with chaotic encryption and compression at the transmitting end and decoding at the receiving end. 
  
\begin{figure}[htbp]\hypertarget{Fig1}{} 
\centerline{\includegraphics[scale =0.45]{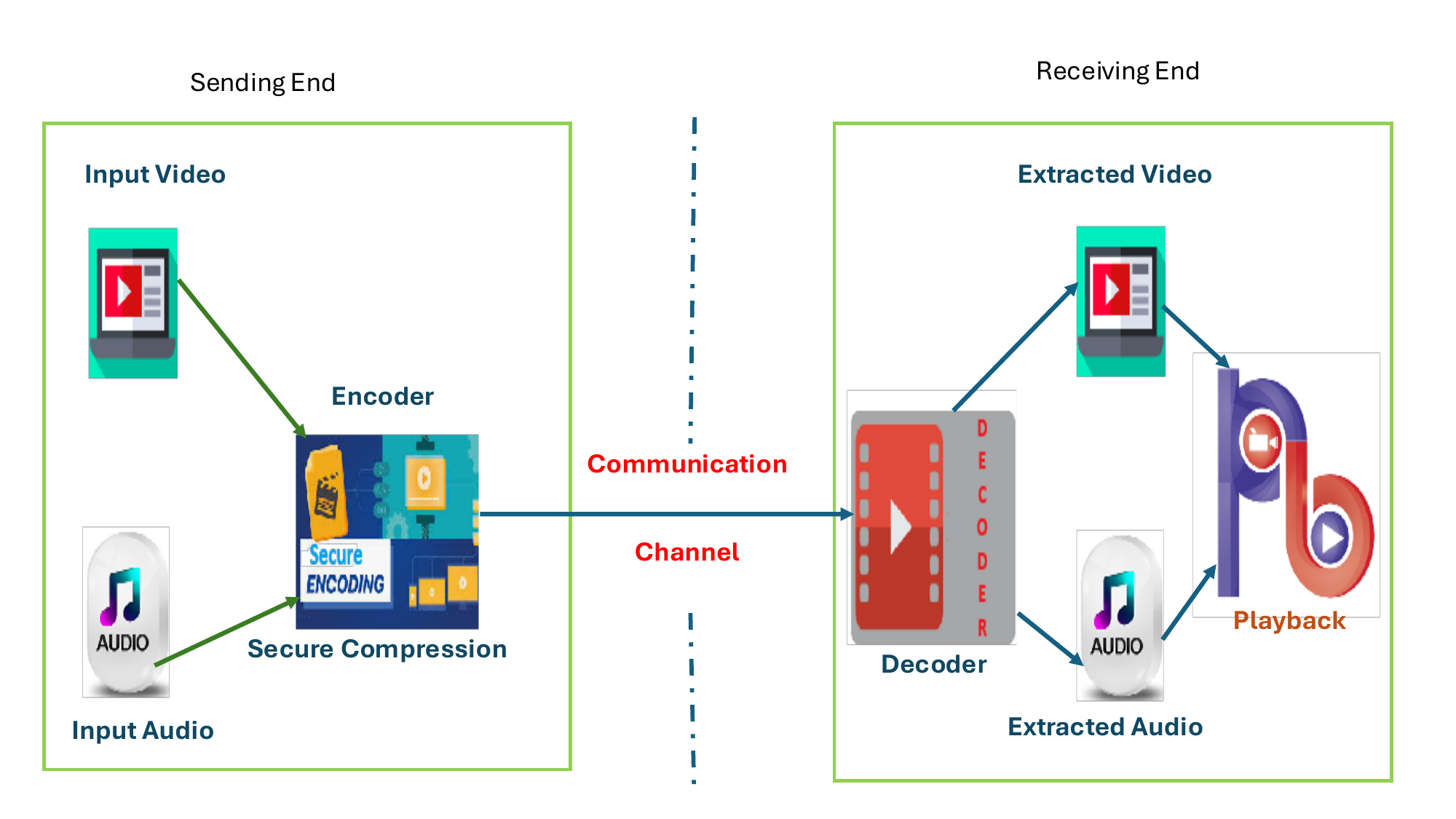}}
\caption {Secure transmission of multimedia data and its retrieval at the terminus}
\end{figure}

\subsection{Challenges}
Maintaining Stoner data sequestration while migrating pall services, distinct operations, the entire pall, or Stoner data transport is relatively gruelling. With an improvement in data size, it becomes further critical. The current literature offers several security mechanisms to protect data sequestration. Among them, data encryption and steganography methods are popular and effective. However, they give advanced time and space complexities and improve information loss \cite{kar2018improved, ali2025novel}. Thus, their operations are not space- and time-complete for large pall data and are hamstrung to save the data integrity. 

Multimedia data such as video and images are critical, as they might contain efficient information. As user and business demands require the creation of better video quality, such as 4K, 8K, and immersive VR/AR content, the need for systems that enable power and security is greater than ever \cite{akkasaligar2020medical,ahuja2021novel}. Ensuring the security of video data storage and transmission while ensuring efficient use is a critical issue. To solve these problems, cryptography and compression techniques are often used together. However, traditional methods of performing compression and encryption as separate operations are increasingly inadequate to meet the scale, complexity, and security needs of today's video systems \cite{wu2024novel}.

\subsection{Problem Statement}
Securing multimedia files requires them to be transported safely. The recipient will receive the data without any loss of information. Often malicious attackers may affect the secure transmission of the file, rendering it completely useless or spreading misinformation \cite{song2019efficient, umar2024chaos}. This research focuses on the following areas: 
\begin{enumerate}[label=(\roman*).]
    \item During the transmission of multimedia files, attackers or third-party invaders can change the information. Smooth file transmission is required, which will maintain data privacy, integrity, and quality \cite{zhang2018lightweight}. It will maintain the privacy between the sender and the receiver and will not leak data outside.
    \item Even though the algorithm is known to the attacker, he must know the confidential data (keys) that are used during encryption and compression \cite{bhattacharjee2023simultaneous, guan2020efficient}. This will only be known to the sender and receiver, and hence this will maintain the privacy of the video file. 
    \item The endeavours to safeguard data secrecy impede data integrity, or conversely \cite{bhattacharjee2019integrated}. The existing literature lacks a unified strategy to tackle confidentiality and integrity issues during extended video transmission over many insecure channels. 
\end{enumerate}

\subsection{Research Objective}
This research on multimedia compression and encryption focuses mainly on performing simultaneous data compression and encryption in a single step. The main objectives of this research are the following.
\begin{enumerate}[label=(\roman*).]
    \item Implementing the algorithm of simultaneous compression and encryption on video files to maintain integrity and security
    \item Since compression and encryption are taking place on a single step, computational burden decreases, and space and time complexity decrease.
    \item Creating a cohesive platform to mitigate confidentiality concerns when delivering huge videos without compromising data integrity. 
\end{enumerate}

\subsection{Research Scope}
The scope of this research is limited to developing an algorithm for video files that will do simultaneous compression and encryption as shown in Fig. 1, which will do both in a single step based on a key. To introduce randomness in the compressed video file and to improve the level of security, a chaotic map is used \cite{darwish2019modified, zia2022survey}. The output video can only be retrieved using the same key.

\subsection{Contribution and Novelty Analysis}
Ensuring the security of substantial multimedia assets during transit is consistently difficult. Diverse factors, including transmission problems, security breaches, environmental noise, and others, significantly diminish data quality, integrity, and privacy in a large multimedia file distribution system \cite{pande2010joint, xu2020robust}. In addition to that, the other main problems with transporting large multimedia files such as video files are insufficient storage and bandwidth. 

The primary contribution of this research is the development of an advanced chaotic approach that mitigates concerns about data privacy and integrity in the transmission of large multimedia files, such as videos, by introducing an innovative chaos-based encryption technology.  This research further contributes as follows.

\begin{enumerate}[label=(\roman*).]
    \item Mitigates many security vulnerabilities, including the prevention of security breaches, the reduction of transmission mistakes, and the minimisation of data loss, among others.
    \item Unique data compression addresses storage challenges at both transmission and reception points, alleviates bandwidth constraints, and further reduces data loss. 
    \item The sophisticated key management system guarantees data privacy and reduces the likelihood of unauthorised tampering by illicit third parties.
\end{enumerate}

\subsection{Research Motivation}
In any extensive video transmission system, safeguarding the integrity and confidentiality of the data is the primary priority.  On several occasions, various security attacks or transmission failures may result in partial or complete corruption of large multimedia assets, such as video files \cite{song2019efficient, pande2010joint}.  In addition, storage and bandwidth constraints represent significant obstacles in extensive video transmission.  However, the current literature does not provide a method to simultaneously handle all these difficulties.  This resource presents an innovative application for choice encryption and compression, which encrypts huge videos during transmission to enhance data privacy while effectively compressing the input video to reduce storage space and bandwidth consumption \cite{bhattacharjee2024leveraging}.  The proposed technique successfully mitigates certain transmission defects without incurring additional space and time overhead.  The literature indicates that data integrity can be sacrificed to improve data privacy or conversely.  The current literature indicates that simultaneously reducing temporal and spatial complications is extremely challenging \cite{zia2022survey}.  Consequently, the suggested selection method has been formulated to address data integrity and privacy concerns simultaneously, without compromising either aspect, while maintaining minimal time and space complications. 

\subsection{Paper Organization}
The structure of this document is as follows: \hyperlink{sec1}{Section 1} provides a concise overview of the security challenges associated with the transmission of large video files and the several innovative strategies to address these challenges.  This section provides a comprehensive explanation of the problem statements, objectives, scope, contribution and novelty analysis, as well as the motivation of this research effort.  Thus, \hyperlink{sec2}{Section 2} provides an exhaustive study to assess the strengths and weaknesses of current security methods to pinpoint the existing research need. In addition, it delineates a comprehensive technique to rectify the current research gap.  \hyperlink{sec3}{Section 3} describes the construction of the proposed technique, and it further discusses the potential use case and analyses the time complexity of it. 
\hyperlink{sec4}{Section 4} outlines the frameworks for experimental setup and data preparation to enable the execution of the suggested technique and its performance evaluations.  \hyperlink{sec5}{Section 5} outlines the necessary parameters and their implementations to illustrate the effectiveness of this contingent strategy in various settings.  The  \hyperlink{sec6}{Section 6} concludes this research and \hyperlink{sec7}{Section 7} foresees the need to improve it in future efforts.  The \hyperlink{appendix}{Appendix} simultaneously presents the sources of the data set used and software applications.

\hypertarget{sec2}{}
\section{Literature Review}

Since the early 1960s, the chaos proposition in nonlinear dynamical systems had been applied in numerous exploration areas because of the capability to respond to effective change to the original conditions and control parameters, which makes the route of the chaotic charts entirely changeable and arbitrary. In fact, numerous experimenters linked close connections between chaos proposition and cryptography that are useful for the practical perpetration of cryptosystems due to exponential parcels of nonlinear dynamical systems similar as ergodicity, perceptivity to the original condition and control parameters. These parcels are directly linked to the confusion and prolixity parcels of classical cryptosystems \cite{karmakar2021sparse, cai2023image}. In addition to the theoretical study of the dynamics of chaos, its implicit operations have also been extensively explored in the development of low-complexity cryptosystems. Likewise, chaotic dynamics is employed in the construction of the S-boxes and design of cryptographic algorithms.
The chaotic sine map, which can be expressed using the following \hyperlink{eqn6}{Equation(6)}, is used in the development of the SBOX.

\hypertarget{eqn6}{\begin{equation} \tag{6}
V_{\omega+1}={\phi}_{\lambda}(V_{\omega})=\lambda \sin(\pi \times V_{\omega})
\end{equation}}
 In \hyperlink{eqn6}{Equation (6)}, $V_{\omega} \in \{0,1\}$, $\omega=\{0,1,2,3,.,.\}$,  $\lambda \in \{0,1\}$ and ${\phi}_{\lambda}(V_{\omega})$ where $\lambda \in \{0,1\}$ are the control parameters in any chaotic sine map \cite{usama2017chaos}. 
Numerous nonlinear dynamical systems can be employed for cryptographic purposes when they are successfully enforced in horizonless perfection computing. Still, the operation of the chaos proposition in the development of contemporaneous data contraction and  encryption systems is rightly studied in comparison with its counter-accusations in cryptography \cite{matin2021video, shi2024heterogeneous}. The chaotic encryption can be achieved with the chaotic logistic map, ${\theta}_{\lambda}({\upsilon}_{\omega})$ where, ${\upsilon}_{\omega} \in \{0,1\}$ and ${\lambda} \in \{0,4\}$ and $\omega=\{1,2,3,4,....,n\}$. It can be formulated with the following \hyperlink{eqn7}{ equation (7)}.  

\hypertarget{eqn7}{\begin{equation} \tag{7}
{\upsilon}_{\omega+1}={\theta}_{\lambda}(\upsilon_{\omega})=\lambda {\upsilon_\omega}(1-{\upsilon}_\omega)
\end{equation}}

In \hyperlink{eqn7}{Equation (7)}, $\upsilon_{0}$ is the initial input video data when $\omega=0$. Chaos, contraction, and cryptography are three separate disciplines, but they've a close relationship with several pros and cons. Compression works by reducing the redundancies present in the data. Cryptographic ways also reduce the redundancies that are set up in the data using the controlled secret key but do not, in general, reduce the data size. Therefore, the main concern of the contraction and cryptography is to reduce the redundancies in the data. \hyperlink{Fig2}{Fig. 2} shows the exploration framework for chaos-based video compression and encryption. These three can be added to compress and cypher data contemporaneously; therefore, it is mainly worth probing chaos, contraction, and cryptography for practical operations \cite{radwan2016symmetric, xu2019commutative}. 

\begin{figure}[htbp]\hypertarget{Fig2}{} 
\centerline{\includegraphics[scale =0.45]{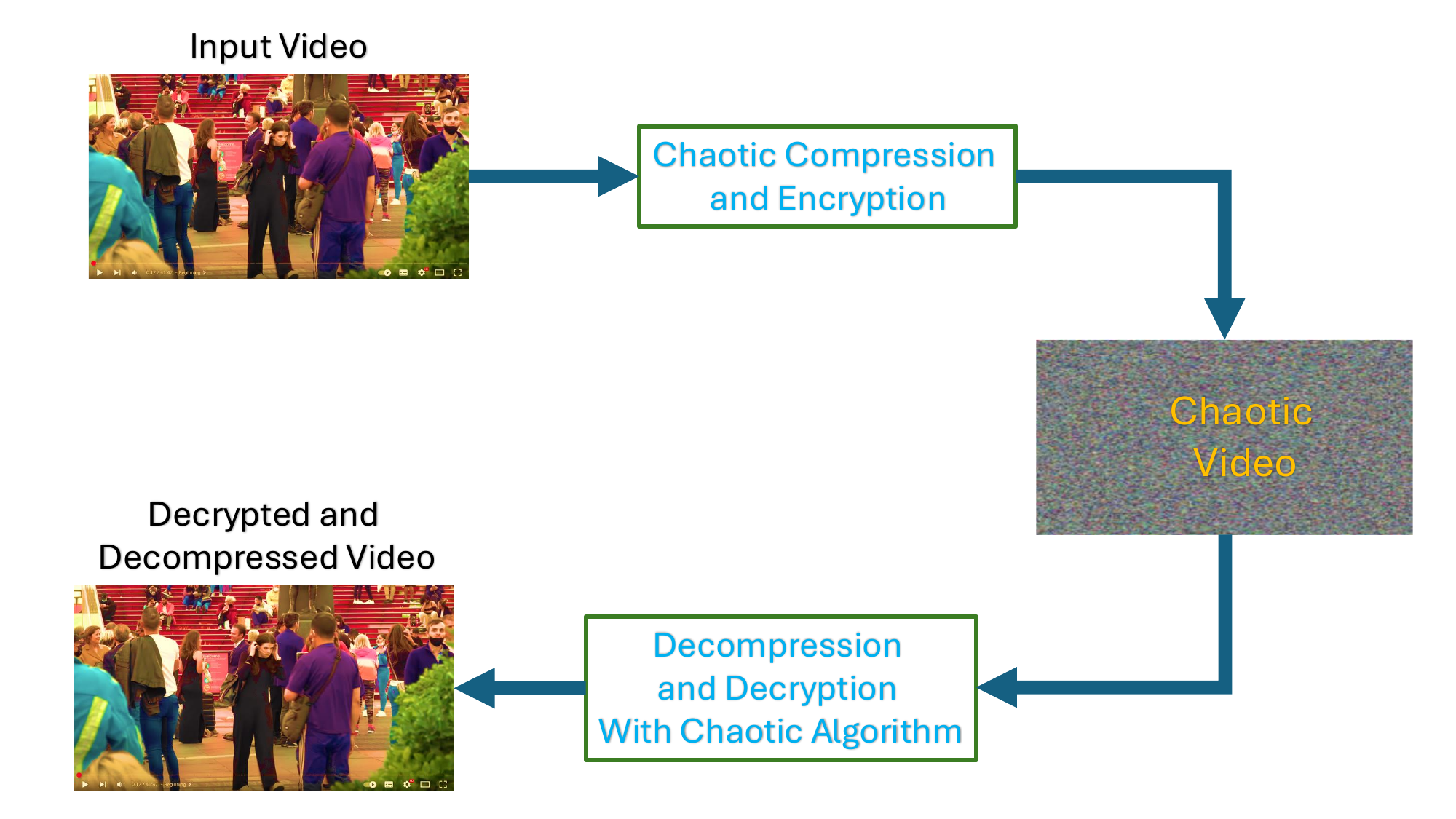}}
\caption {Chaotic encoding of video data and its retrieval}
\end{figure}

\subsection{Data Compression}
Data compression reduces data redundancy by converting it into a form that requires less space to save space and reduce transmission time, especially when the initial data has more redundancy to avoid overhead and reduce network utilisation \cite{zhu2019high,lin2019novel}. Data compression techniques can be of two types: lossy and lossless. Lossy compression continuously deletes unnecessary and unimportant data during the compression process, thereby reducing data quality. \hyperlink{Fig1}{Fig. 1} shows the lossy compression algorithm. JPEG, MPEG, TIF, PING, and MP4 are good examples of lossy compression \cite{bhattacharjee2023simultaneous}. Any compression technique's ability to yield a higher compression percentage can be used to measure its effectiveness  \cite{bhattacharjee2015lossless,bhattacharjee2024leveraging} . The compression percentage ($\Omega$) can be computed using the following \hyperlink{eqn8}{Equation (8)} if the original file size is represented by $\Theta$ and the compressed data size in any data compression method is represented by $\Gamma$. 

\hypertarget{eqn8}{\begin{equation} \tag{8}
\Omega=(1-\frac{\Gamma}{\Theta}) \times 100
\end{equation}}
However, lossless compression such as RLE, Shannon-Fano, HC, AHC, AC, and LZ77 provides data similar to the original. Lossless compression does not remove the data as each data bit remains the same after decompression, hence maintaining the data quality after decompression \cite{bhattacharjee2015lossless, long2025improved}. This is a more efficient way of compression when data integrity and quality is concerned.Compression does not employ secret key or word restrictions, making the compressed file vulnerable to unauthorised access. Thus, when compromised data are stored on bias or transferred over communication channels, they remain vulnerable to unauthorised access and illegal use. Currently, video security is a major challenge because video files may contain confidential data. Therefore, to apply cryptography, it is critical to keep the data storage and transmission system secure \cite{ali2025novel, akkasaligar2020medical}. One of the problems with any chaos-based security system is data loss. The research demonstrates that there are several reasons why any secure video transmission system could experience data loss. Any security system's stability, which is determined by its ability to generate a larger entropy value, can determine how strong it is. The efficiency of any secure transmission method can be determined by the category entropy loss due to the various causes of data loss. An inefficient system is indicated by a greater category entropy loss score, or vice versa \cite{umar2024chaos}. The following \hyperlink{eqn9}{Equation (9)} can be used to calculate the categorical entropy loss ($\xi$) for any transmitted video data in any chaos-based secure data transmission system if the transmitted data is represented by $\gamma$ and the retrieved data at the receiving end is represented by $\hat{\gamma}$. \hyperlink{eqn9}{Equation (9)} is derived as follows when the number of input video samples is denoted as  $\eta$ and the number of iterations is represented by $\sigma$.

\hypertarget{eqn9}{\begin{equation} \tag{9}
\xi=-\sum_{i=1}^{\eta}\sum_{j=1}^{\sigma} {\gamma}_{ij}\log_{10}{\hat{\gamma}_{ij}} 
\end{equation}}

The workings of lossy and lossless compression are illustrated in \hyperlink{Fig3}{Fig.3}. This exploration work focuses substantially on developing a contemporary compression and encryption fashion that would not compromise data quality. Hence, lossless condensing serves the purposes. Therefore, the discussion of lossy compression is neither considered nor included in the following discussions \cite{cai2023image, camtepe2021compcrypt}.

\begin{figure}[htbp]\hypertarget{Fig3}{} 
\centerline{\includegraphics[scale =0.45]{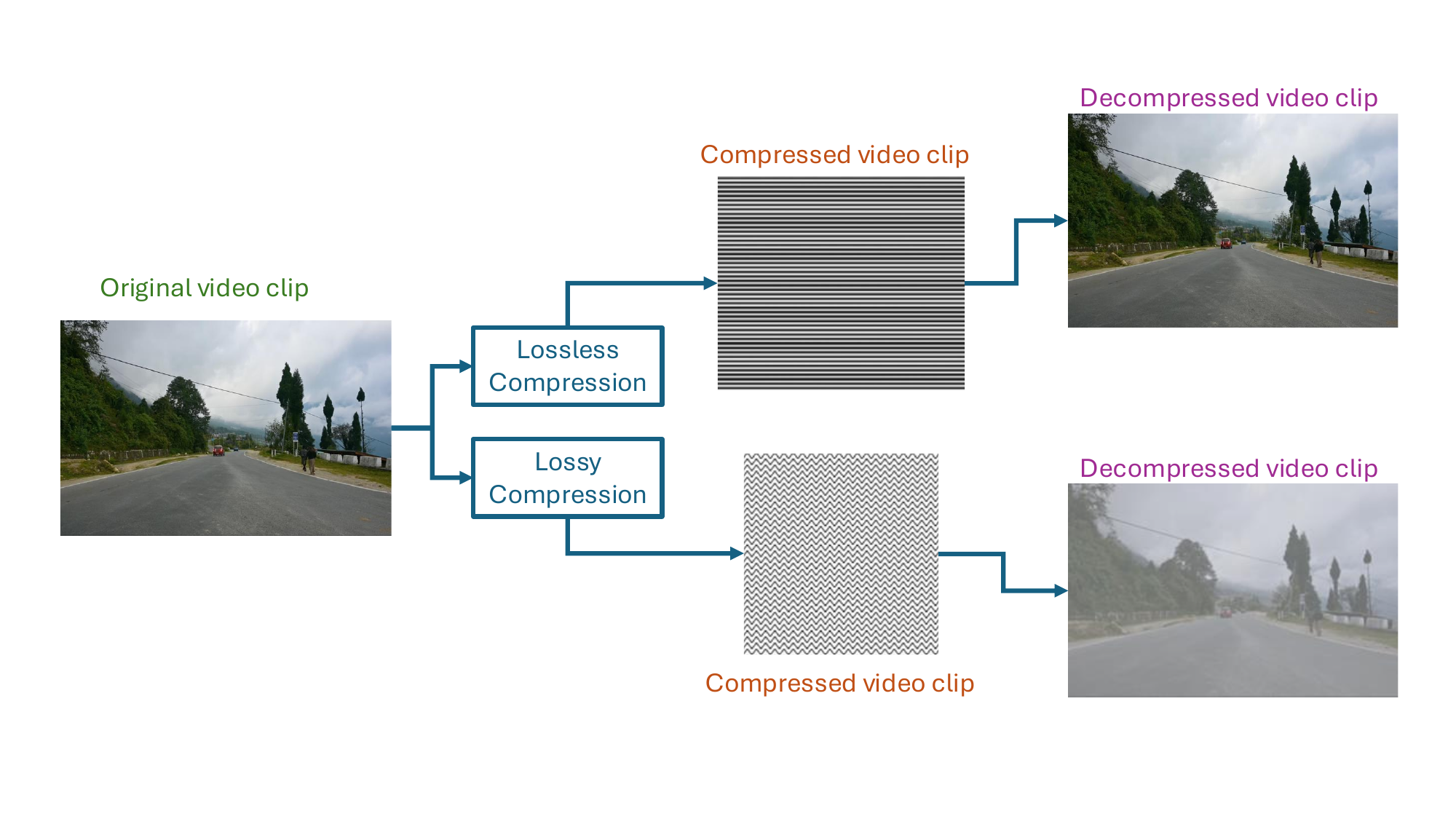}}
\caption {Workings of lossless and lossy compression}
\end{figure}

As depicted in \hyperlink{Fig3}{Fig. 3}, lossy compression involves the permanent removal of certain statistically redundant data sets to enable data compression.  The complete removal of redundant data sets during compression results in outstanding compression efficiency. In scenarios where the preservation of high data quality is not critical, lossy compression is preferable for these applications. The primary lossy compression methods for audio, video, image, and text media files include MP3, MP4, MPEG, and JPEG \cite{singh2021level, chai2020efficient}.  Consequently, lossy compressions are inadequate for research purposes. In lossless compression, no irreversible data loss occurs, as each data bit remains identical to the original after decompression. Articles \cite{bhattacharjee2015lossless, long2025improved} assert that with lossless compression approaches, data are reorganised within the file for improved efficiency. Due to the absence of data loss in lossless compression, the size of a losslessly compressed file exceeds that of a lossy compressed file \cite{umar2024chaos}. A comparative analysis is further shown in \hyperlink{table1}{Table 1} in terms of their strengths and weaknesses. 

\setlength{\tabcolsep}{2pt}        
\renewcommand{\arraystretch}{1.08} 
\setlength{\leftmargini}{1.2em}  

\begin{table}[htbp]
\centering
\footnotesize
\hypertarget{table1}{}
\caption{Performance analyses of different data compressions}

\begin{tabular}{|p{2.4cm}|p{4.3cm}|p{4.3cm}|p{2.4cm}|}
\hline
\multicolumn{1}{|c|}{\textbf{Approaches}} &
\multicolumn{1}{c|}{\textbf{Strengths}} &
\multicolumn{1}{c|}{\textbf{Weaknesses}} &
\multicolumn{1}{c|}{\textbf{Discussions}} \\
\hline

Molecular bio-inspired chaotic video compression
\cite{kar2018improved} \cite{shanmugam2025optimizing} \cite{karmakar2021sparse}
\cite{bayari2025novel} \cite{zhu2019high} \cite{gomes2025end}
\cite{liu2025secure} \cite{sharma2025quantum}. &
\begin{enumerate}[leftmargin=*]
  \item The roll-out of context-adaptive coding provides enhanced confidentiality with reduced computational cost.
  \item The use of multidimensional chaotic map-based encryption and compression improves data security and portability.
\end{enumerate}
&
\begin{enumerate}[leftmargin=*]
  \item Inadequate assessment of frame rate distortion might affect video quality.
  \item The insufficient temporal compression and lack of an effective dictionary resulted in a diminished compression ratio.
\end{enumerate}
&
An advanced chaotic framework is necessary for managing multi-modal data and ensuring requisite security and portability. \\
\hline

Chaotic logistic-map-based H.264 advanced video coding
\cite{tabash2019efficient} \cite{song2019efficient} \cite{sethi2022joint}
\cite{xu2020robust} \cite{shao2024multi} \cite{singh2021level}
\cite{darwish2019modified} \cite{singh2024systematic}. &
\begin{enumerate}[leftmargin=*]
  \item Introduction of the hybrid chaotic map for exposition-enhanced data security and portability through increased complexity of behaviour.
  \item Multi-objective optimisation of video data using a quantum chaotic map and sparse coding to meet dominant time-varying constraints for video streaming, preserve video quality, and minimise bandwidth.
\end{enumerate}
&
\begin{enumerate}[leftmargin=*]
  \item Rising challenges include the lack of an effective hybrid chaotic solution, video processing complexity and unpredictability, AI-driven cryptoanalysis, and quantum computation malfunctions.
  \item The current hybrid chaotic system is too complex and power-hungry for advanced meta-verse and augmented reality applications.
\end{enumerate}
&
The enhancement of current chaotic systems is crucial to address numerous advanced difficulties, whether arising from AI-driven methodologies or failures in quantum computing characterised by low power consumption and complexities. \\
\hline

Multi-dimensional chaos-based multimedia compression for distributed computation roles
\cite{preishuber2018depreciating} \cite{long2025improved} \cite{shanmugam2025optimizing}
\cite{van2013encryption} \cite{salunke2021beta} \cite{cai2023image}
\cite{okunbor2025analysis} \cite{zhu2024visual} \cite{chai2020efficient}
\cite{gao2025enhanced} \cite{liu2025enhancing} \cite{li2025error}
\cite{mahalakshmi2025comprehensive}. &
\begin{enumerate}[leftmargin=*]
  \item Utilisation of a multi-key chaotic map using symmetric numeral systems and Fourier-based cosine transformation to enhance compression ratios and video security.
  \item An elementary cellular automata-based block compressive four-dimensional chaotic coding map improves video transmission portability, security, and robustness.
\end{enumerate}
&
\begin{enumerate}[leftmargin=*]
  \item The frame distortion problems and increased complexity diminish video quality, integrity, and robustness.
  \item The elevated temporal complexity associated with generating chaotic and cryptic real-time embedded video exacerbates data loss and diminishes data integrity.
\end{enumerate}
&
An efficient and lightweight chaotic compression method is necessary for big video transmission that maintains acceptable video quality, robustness, and data integrity. \\
\hline

Commutative chaotic code-based compression to transfer large video over heterogeneous networks for the various industrial applications and usage
\cite{matin2021video} \cite{bhattacharjee2024leveraging} \cite{bhattacharjee2021study} \cite{bhattacharjee2015lossless} \cite{bhattacharjee2023simultaneous}
\cite{pande2010joint} \cite{xu2019commutative} \cite{guan2020efficient} \cite{lin2019novel} \cite{esakki2021adaptive} \cite{chen2025security} \cite{lee2020start}. &
\begin{enumerate}[leftmargin=*]
  \item Uses of symmetric permutation matrices and fractal skewed chaotic maps to enhance security and portability of video data in the cloud.
  \item Data integrity and privacy improved with CLM and SBOX-based multi-dimensional chaotic-map.
\end{enumerate}
&
\begin{enumerate}[leftmargin=*]
  \item Insufficient to provide adequate anti-noise performance to ensure robustness under diverse channel circumstances.
  \item The increased complexity of the employed neural network-based chaotic compression methods undermines data integrity.
\end{enumerate}
&
A proficient and resilient chaotic compression method is necessary for transmitting huge videos with minimal complexity. \\
\hline

\end{tabular}
\end{table}

\hyperlink{table1}{Table 1} indicates that the existing chaotic compression-based methods inadequately provide the necessary robustness, integrity, and confidentiality simultaneously. Certain procedures do not ensure the preservation of video quality either. Time and space complications present additional challenges in video data processing using several existing chaos-based methodologies. These difficulties escalate as the video size increases. Consequently, the research in \hyperlink{table1}{Table 1} demonstrate that the existing literature does not provide any sufficient chaotic methodologies that can address.

\subsubsection{Chaos Theory}
Chaos theory has its origins in weak systems, known for their unpredictable yet stable behavior, often described as the "butterfly effect". This sensitivity to initial conditions and the controllability of chaotic systems is an attractive tool to improve the security and efficiency of information processing. By applying the principles of ergodicity, unpredictability, and sensitivity to initial state, chaos can provide effective solutions for cryptography and compression \cite{okunbor2025analysis,wen2023chaos}. It is related to the intended use of cryptocurrency. Obfuscation ensures that the relationship between the cipher text and the encryption key is as complex as possible, while diffusion makes it difficult to identify the bad analysis by revealing the consequences of a single bit in the cipher text \cite{xu2020robust}. The idea of employing chaos to design cryptographic systems can be found in the classic Shannon masterclass research work entitled 'Communication Theory of Secrecy Systems', published in 1949: 'Good mixing transformations are often formed by repeated products of two simple non-commuting executions. Let  the initial state  of a chaotic map $(f)$ be the $\alpha_0$ and the divergence of input data after a duration $(\lambda)$  and $\eta$ iterations is $(\alpha_0 + \varepsilon)$ where $\eta>0$.  The characteristic equation \cite{schuster1988deterministic} of applied chaotic approach can be formulated with the following \hyperlink{eqn10}{Equation (10)}.   

\hypertarget{eqn10}{\begin{equation} \tag{10}
|f^{\eta}({\alpha_0} + \varepsilon)-f^{\eta}(\alpha_0)|=\varepsilon \times e^{\eta} \times \lambda(\alpha_0)
\end{equation}}

It has shown, for example, that pastry dough can be mixed using such a sequence of operations. The dough is first rolled onto a thin slab, then folded over, then rolled, and then folded again. Chaotic maps, including imaginary maps, sine maps \cite{kumari2024novel}, and Arnold cat maps, form the basis for creating pseudorandom keystream generators and key boxes, which are important components of encryption schemes \cite{singh2024systematic, muthu2021review}. 

\subsection{Cryptography}
 Encryption ways are classified into symmetric- key( private- key) and asymmetric- key( public- key) cryptography. Symmetric- crucial cryptography uses the same key for encryption and decryption, while asymmetric- crucial cryptography employs a public-private crucial brace. Asymmetric systems, such as RSA, have mechanisms for crucial generation or operation. Symmetric styles, similar to DES, AES, and Triple DES, are generally used and can be distributed into block ciphers (cracking fixed-size data blocks) and sluice ciphers (processing data bit- by-bit or byte-by-byte). 
 Block ciphers, such as AES, provide advanced security at the cost of complexity, while sluice ciphers, such as RC4, offer brisk performance. Strong encryption generally involves multiple duplications and trading speed for security. Inventions, similar to S-Box optimisation and GPU-accelerated fabrics, aim to ameliorate performance. Nevertheless, these S-Box-based methods for creating block ciphers don't always produce the intended outcome. If $\zeta_{in}$ represents the input differences and $\zeta_{out}$ represents the output differences, as the S-Box-based block ciphertext generation technique resulted from a number of system defects. The following \hyperlink{eqn11}{Equation (11)} can be used to do the differential fault analysis \cite{chai2020efficient, wen2023chaos,hosny2023fast}.
 
 \hypertarget{eqn11}{\begin{equation} \tag{11}
S(\alpha \oplus \zeta_{in}) \oplus S(\alpha)=\zeta_{out}
\end{equation}}
 
 In \hyperlink{equn11}{Equation (11)}, $S(\alpha)$ denotes the used S-Box mechanism where, $\alpha$ is a dependent sub-key.  Cryptography relies on defying cryptanalysis, which seeks to expose secured data using methods like algebra and the number proposition. Common attacks include ciphertext- only, known-plaintext, chosen-plaintext, chosen-ciphertext, and brute- force styles. Effective cryptographic systems carry large crucial sizes, arbitrary labors, and perceptivity to input changes to repel these attacks. In any  security approach the attacker primarily tries to find the security key $(\kappa)$ within the following \hyperlink{eqn12}{Equation (12)}. 
\hypertarget{eqn12}{}
\begin{equation}\tag{12}
    \begin{cases}
      {\xi}_1=\partial_{\kappa}({\rho}_1)\\
      {\xi}_2=\partial_{\kappa}({\rho}_2)\\
      \vdots \\
       {\xi}_n=\partial_{\kappa}({\rho}_n)\\
    \end{cases}       
\end{equation}

 In \hyperlink{eqn12}{Equation (12)}, $n$ the number of rounds in which the input samples $(\rho_n)$ are subjected to the cryptic technique $(\partial_k)$. The act of ensuring data security for storage or communication. It should be replaced in such a way that it becomes delicate for an unethical existent or group to be able to determine its true essence. Compact in freeing the computer and securing it safely, several technologies come as close as encryption to nearly unbreakable information security, such as getting the data and executing it within an encryption algorithm, and it is undecipherable to everyone who does not maintain the right decryption process to convert the system. The biases that are employed to cipher the data must be streamlined or use streamlined software and be free of any defects or bugs so that hackers are not able to take advantage of any vulnerability of the system to get access to the information. 

\subsubsection{Pseudorandom Keystream Generator}
A pseudo-random keystream generator, or PRKG is a critical element in cryptography used to produce a sequence of bits or bytes that appear arbitrary but are generated deterministically using a specified key. The generated keystream is used to mask plaintext during encryption, generally through operations such as XOR or XNOR. Chaos-ground PRKGs influence the essential unpredictability and perceptivity of chaotic systems to cosign conditions, making them largely suitable for secure encryption. If the probability of any input video samples $(\theta)$ is represented by $\{ {\sigma}_1, {\sigma}_2, {\sigma}_3,....,{\sigma}_n \}$ in a pseudo-random key-stream generation  \cite{luca2004new} for any chaotic video encryption scheme, and if the employed method is represented by $\oint(\theta)$, we obtain the following \hyperlink{eqn13}{Equation (13)} and \hyperlink{eqn14}{Equation (14)}. 

\hypertarget{eqn13}{}
\begin{equation}\tag{13}
   \oint(\theta)= \begin{cases}
      \frac{\theta}{\sigma_1} \ where, \ \theta \in \gamma_1\\
      \frac{\theta-\sigma_1}{\sigma_2} \ where, \ \theta \in \gamma_2\\
      \vdots \\
    \frac{\theta-\sum_{i=1}^{n-1}{\sigma_i}}{\sigma_n} \ where, \ \theta \in \gamma_n\\
    \end{cases}       
\end{equation}

\hypertarget{eqn14}{}
\begin{equation}\tag{14}
   \gamma_i =\left[\sum_{j-1}^{i-1}{\sigma_j} \ , \sum_{j=1}^{i}{\sigma_j}\right] \ where, i=\{2,3,...,n\}
\end{equation}

By using chaotic charts similar to the logistic chart, tent map, or Arnold cat map, these creators produce keystreams with strong randomness and unpredictability. However, achieving a balance between randomness, effectiveness, and computational budget remains a crucial challenge in PRKG design. 

\subsubsection{Chaotic Logistic Map}
The logistic map is a simple yet important one-dimensional chaotic chart extensively used in cryptography due to its capability to induce pseudorandom sequences. The initial state of a chaotic logistic map \cite{zhang2005image} $\lambda(\nu)$ is represented by $\nu_0$. The other state can be computed using the following \hyperlink{eqn15}{Equation (15)}.

\hypertarget{eqn15}{}
\begin{equation}\tag{15}
   \begin{cases}
   \nu_1=\lambda(\nu_0) \\
   \nu_2=\lambda(\lambda(\nu_0))\\
   \nu_3=\lambda(\lambda(\lambda(\nu_0))) \\
    \vdots  \\
    \nu_n=\lambda(\lambda(\lambda(\hdots \hdots \lambda(\nu_0))))
   \end{cases}
\end{equation}
This perceptivity to original conditions and control parameters makes the logistic chart ideal for cryptographic operations, similar to pseudorandom keystream generation, where unpredictability and randomness are essential. It is computationally effective and easy to apply, making it a popular choice for featherweight cryptographic systems. Still, it can face challenges like non-uniform distribution and amiss chaotic ranges, which bear careful tuning to enhance security and effectiveness in encryption schemes. 

\subsubsection{Chaos-Based Secure Video Compression}
Huffman compression is a well- known data contraction fashion. It consists of two main corridors, that is, the statistical model known as the Huffman tree and the compression part. It utilises the Huffman tree to law the data symbols by assigning them shorter devices. However, to negotiate the contraction process, it requires scanning of the input data twice. First, it needs to produce a statistical model from input data symbols to construct the Huffman tree. Second, Compression machine assigns a shorter metaphor to input data symbols using a constructed Huffman tree \cite{hermassi2010joint}. The mean code length $(\gamma)$ for any compression method based on Huffman coding can be determined using \hyperlink{eqn16}{Equation (16)}, and with the help of \hyperlink{eqn16}{Equation (16)}, we can calculate the total number of extracted bits after the decompression. The total number of extracted bits can be calculated with the following \hyperlink{eqn21}{Equation (21)}.

\hypertarget{eqn16}{}
\begin{equation}\tag{16}
\gamma=\frac{\sum_{i=1}^{n}{(\theta_i \times \zeta_i)}}{\sum_{i=1}^{n}{(\theta_i)}}
\end{equation}

In \hyperlink{eqn16}{Equation (16)}, $\theta_i$ denotes the frequencies of each distinct input character, while $\zeta_i$ signifies the length of each Huffman compressed code associated with a specific character, where $n$ indicates the total number of unique characters.  An improved interpretation of the Huffman compression was presented in that does not require surveying two times for data contraction, known as Adaptive Huffman Coding or AHC \cite{usama2021efficient}. AHC also consists of two main corridors, i.e., the statistical model and the contraction machine, but it mutates the statistical model known as Adaptive Huffman Tree (AHT) and assigns the shorter devices at the same time. Therefore, it needs to overlook the input data only once to complete the contraction process. AHT plays an essential part in AHC rendering process to achieve compression effectively (when AHT matches the exact statistical characteristics of data). More importantly, there are three advantages to these two separate corridors that are used as design principles to apply the proposed work. 

\begin{enumerate}[label=(\roman*).]
     \item AHC allows equivocation or randomisation of symbols chances during the AHT generation process that leads to the achievement of Shannon’s suggested abecedarian parcels of confusion and prolixity.
     \item AHC allows symbol negotiation or changes the order of symbols before updating AHT and data garbling, giving significant inflexibility to introduce complications in the contraction process without compromising contraction capabilities.
     \item AHC allows integrating different cryptographic processes that enable masking the rendering affair with some pseudorandom keystream that can enhance encryption quality.
 \end{enumerate}

The existing literature demonstrates the application of numerous chaotic encryption methods and separate chaotic map-based strategies to assure data security in the transmission of diverse video and image data. A concise analysis of their strengths and weaknesses has been elaborated in \hyperlink{table2}{Table 2}.  Additionally, presents a general comment based on the strengths and weaknesses of existing approaches.

\setlength{\tabcolsep}{2pt}        
\renewcommand{\arraystretch}{1.08} 
\setlength{\leftmargini}{1.2em}  

\begin{table}[htbp]
\centering
\footnotesize
\hypertarget{table2}{}
\caption{Performance analyses of different chaotic encryptions}

\begin{tabular}{|p{2.4cm}|p{4.3cm}|p{4.3cm}|p{2.4cm}|}
\hline
\multicolumn{1}{|c|}{\textbf{Approaches}} &
\multicolumn{1}{c|}{\textbf{Strengths}} &
\multicolumn{1}{c|}{\textbf{Weaknesses}} &
\multicolumn{1}{c|}{\textbf{Discussions}} \\
\hline
A lightweight multi-key generation and chaotic coding-based video/image encryption \cite{huang2025efficient} \cite{long2025improved} \cite{zhang2018lightweight} \cite{van2013encryption} \cite{shifa2020mulvis} \cite{hosny2023fast} \cite{kumar2025image}. &
\begin{enumerate}[leftmargin=*]
    \item Employment of a lightweight chaotic map-based video encryption method for secure transmission inside a specific framework.
    \item Utilisation of very efficient coding for partial video encryption to improve portability and security in diverse real-time applications.
\end{enumerate}
&
\begin{enumerate}[leftmargin=*]
    \item Inadequately safeguards against a diverse array of security threats and faults in video data transmission.
    \item Does not safeguard against the extensive security threats against quality and errors in video transmission.
\end{enumerate}
&
An enhanced chaotic structure is necessary to address the diverse array of security threats and transmission failures across many application domains.
\\
 \hline
Multimodal chaotic-map based superencryption  for securing communication with image/video \cite{wu2024novel} \cite{el2023even} \cite{ayubi2021new} \cite{kumari2024novel} \cite{muthu2021review} \cite{zia2022survey} \cite{bayari2025novel} \cite{tiwari2025compressed} \cite{lu2025novel} \cite{verma2026optical} \cite{lai2026fast}. &
\begin{enumerate}[leftmargin=*]
    \item A coding-based chaotic map is suggested to provide protection against diverse brute force and statistical attacks.
    \item  The quantum chaotic map is employed in the spatial realm to encrypt image/video pixel blocks, safeguarding against various plaintext attacks by introducing a distinctive context key.
\end{enumerate}
&
\begin{enumerate}[leftmargin=*]
    \item The prevailing quantum-based chaotic algorithms do not provide sufficient security for extensive video transmission across many optical applications.
    \item The current chaotic encryption technique fails to address specific video processing assaults and video recording vulnerabilities, and data integrity is compromised due to insufficient encryption speed.
\end{enumerate}
&
A more robust chaotic encryption methodology is required to manage the complexity of extensive video data processing and to address many possible security problems. 
\\
\hline
Cross-functional applications of multiscroll memristive chaotic compressive encryption \cite{shi2024heterogeneous} \cite{radwan2016symmetric} \cite{kaccar20224d} \cite{wen2023chaos} \cite{karimi2025bi} \cite{ahuja2021novel} \cite{gao2025chaos} \cite{guo2025novel} \cite{lai2026design}. &
\begin{enumerate} [leftmargin=*]
    \item A multi-dimensional chaos mapping is presented in conjunction with the fractional cosine transformation to enhance the security of images and videos.
    \item An effective logistic chaotic tent map with chaotic digital separation loops is introduced for privacy protection and enhancement of video quality.
\end{enumerate}
&
\begin{enumerate} [leftmargin=*]
\item Existing chaotic encryption techniques cannot improve multimedia protection, retain quality, or secure multi-face video/images.
\item The fragility or incapacity of current chaotic solutions to prevent unauthorised access to transmitted large video compromises data security and integrity of any multimedia communication system.
\end{enumerate}
&
The current system needs a solid method that can guarantee the security, integrity, and actual quality of transmitted video while granting authorised access. 
\\ 
\hline
Lightweight visual encryption employing IoT to augment the security and integrity of extensive data transfers across diverse smart city applications. \cite{tabrizchi2020survey} \cite{bhattacharjee2014multibit} \cite{bhattacharjee2019integrated} \cite{camtepe2021compcrypt} 
\cite{bhattacharjee2016security} \cite{ahuja2023iot} \cite{nazish2025cm} \cite{ali2025novel} \cite{peng2026injecting}. &
\begin{enumerate}[leftmargin=*]
    \item Asymmetric numeral pseudorandom bit generator-based chaotic encryption is employed to secure video data transmission.
    \item These approaches employ elliptic curve cryptography for IoT devices to guarantee protection against fault injection attacks with minimal time complexity.
\end{enumerate}
&
\begin{enumerate}[leftmargin=*]
    \item Pseudo-random chaotic encryptions are insufficiently effective to offer strong security features for real-time multimedia and to protect against a variety of new threats.
    \item The chaotic adaptive and electric curve-based cryptographies fail  to offer acceptable security in image/video transmission.
\end{enumerate}
&
More improvements are necessary in the pseudo-random chaotic encryption to ensure sufficient security and integrity for extensive image and video communications.\\
 \hline
\end{tabular}
\end{table}

\hyperlink{table2}{Table 2} examines the advantages and disadvantages of different chaotic encryption methods currently in use for maintaining data confidentiality and integrity when processing massive amounts of video data. It demonstrates that the existing chaotic encryption techniques are unable to provide sufficient time and space efficiency in addition to being unable to handle a variety of transmission failures. Channel congestion is one of the major issues with massive video transmission. This problem gets more crucial because of the high time and space complexity. It can result in data loss and compromise data integrity. In addition, ineffective chaotic encryption may reduce the confidentiality of the data. Therefore, handling all of these situations at the same time requires some work.

The subsequent table, \hyperlink{table3}{Table 3}, analyses the efficacy of diverse chaotic combinations of various available encryption and compression methods in the transmission of huge video data. It delineates the strengths and shortcomings of many existing methodologies in providing data integrity and secrecy, resilience, and the ability to safeguard video quality, categorised by their publication year. The investigation in \hyperlink{table3}{Table 3} identifies the existing literature gap in large video processing and transmission.
\begin{center}
\begin{table}
\hypertarget{table3}{}
\caption{Summary of literature by year to demonstrate their ability to mitigate video data transmission challenges}
\begin{tabular}{|p{2.1cm}|p{1.7cm}|p{1.9cm}|p{2.2cm}|p{2.1cm}|p{2.4cm}|p{1.7cm}|}
\midrule
\centerline{Publication } &\centerline{Articles} & \centerline{Offer Time}& \centerline{Resolve } & \centerline{Offer } &  \centerline{Resolve} &  \centerline{Maintain  } \\
\centerline{Year} & & \centerline{and space} & \centerline{robustness} &\centerline{compression} &  \centerline{confidentiality} &  \centerline{video}   \\
&  & \centerline{efficiency} &  \centerline{and integrity} &  \centerline{efficiency}  & \centerline{issues} &  \centerline{quality} \\
&  &  &  \centerline{issues} &   &  &   \\
\hline
\centerline{2018} &
\centering\cite{kar2018improved}, \cite{tabash2019efficient}, \cite{preishuber2018depreciating},
\cite{song2019efficient},
\cite{chai2020efficient}

&  \vspace{0.05cm}\centerline{\faDotCircleO} 
& \vspace{0.05cm}\centerline{\faCircle}
& \vspace{0.05cm}\centerline{\faCircle}
& \vspace{0.05cm}\centerline{\faCircle}
&  \vspace{0.05cm}\centerline{\faCircleO} \\
\hline
\centerline{2019} & 
\centering \cite{zhu2019high}, \cite{lin2019novel}, \cite{xu2020robust}, \cite{darwish2019modified}

& \vspace{0.05cm}\centerline{\faDotCircleO} 
& \vspace{0.05cm}\centerline{\faCircle}
& \vspace{0.05cm}\centerline{\faCircle}
&\vspace{0.05cm}\centerline{\faDotCircleO} 
& \vspace{0.05cm}\centerline{\faDotCircleO}  \\
\hline
\centerline{2020} &
\centering \cite{shifa2020mulvis}, \cite{lee2020start}, \cite{bhattacharjee2019integrated}, \cite{bhattacharjee2021study}, \cite{tabrizchi2020survey}, \cite{guan2020efficient}, \cite{ayubi2021new}
& \vspace{0.05cm}\centerline{\faCircle}
& \vspace{0.05cm}\centerline{\faDotCircleO}
& \vspace{0.05cm}\centerline{\faCircleO}
& \vspace{0.05cm}\centerline{\faCircle}
& \vspace{0.05cm}\centerline{\faCircleO} \\
\hline
\centerline{2021} & 
\centering \cite{matin2021video}, \cite{esakki2021adaptive}, \cite{karmakar2021sparse}, \cite{salunke2021beta}, \cite{camtepe2021compcrypt}, 
\cite{singh2021level},
\cite{muthu2021review}
& \vspace{0.05cm}\centerline{\faDotCircleO}
& \vspace{0.05cm}\centerline{\faDotCircleO}
&\vspace{0.05cm}\centerline{\faDotCircleO}
& \vspace{0.05cm}\centerline{\faCircle}
& \vspace{0.05cm}\centerline{\faCircleO}\\
\hline
\centerline{2022} &
\centering \cite{kaccar20224d}, \cite{cai2023image}, \cite{hosny2023fast}, \cite{zia2022survey}, \cite{el2022chaos}, \cite{sethi2022joint}
& \vspace{0.05cm}\centerline{\faDotCircleO}
& \vspace{0.05cm}\centerline{\faCircle}
& \vspace{0.05cm}\centerline{\faDotCircleO}
& \vspace{0.05cm}\centerline{\faCircle}
& \vspace{0.05cm}\centerline{\faCircleO} \\
\hline
\centerline{2023} & 
\centering \cite{shi2024heterogeneous}, \cite{el2023even}, \cite{wen2023chaos}, \cite{bhattacharjee2023simultaneous}, \cite{ahuja2023iot}
& \vspace{0.05cm}\centerline{\faCircleO}
& \vspace{0.05cm}\centerline{\faDotCircleO}
& \vspace{0.05cm}\centerline{\faDotCircleO}
& \vspace{0.05cm}\centerline{\faDotCircleO}
& \vspace{0.05cm}\centerline{\faDotCircleO} \\
\hline
\centerline{2024} &
\centering \cite{singh2024systematic}, \cite{gong2024exploiting}, \cite{shao2024multi}, \cite{zhu2024visual}, \cite{akkasaligar2020medical}, \cite{kumari2024novel}, \cite{umar2024chaos}, \cite{long2025improved}, \cite{wu2024novel} 
& \vspace{0.05cm}\centerline{\faCircle}
& \vspace{0.05cm}\centerline{\faCircle}
& \vspace{0.05cm}\centerline{\faDotCircleO}
& \vspace{0.05cm}\centerline{\faDotCircleO}
& \vspace{0.05cm}\centerline{\faDotCircleO} \\
\hline
\centerline{2025} & 
\centering \cite{karimi2025bi}, \cite{okunbor2025analysis}, \cite{gao2025enhanced}, \cite{gao2025chaos}, \cite{nazish2025cm}, \cite{kumar2025image}, \cite{kouadra2025new}, \cite{gao2025encrypt}, \cite{bayari2025novel}, \cite{ali2025novel}, \cite{liu2025enhancing}, \cite{tiwari2025compressed}, \cite{li2025error}, 
\cite{mahalakshmi2025comprehensive}, \cite{guo2025novel}, \cite{ding2025deepface}, \cite{liu2025secure}, \cite{lu2025novel}, \cite{sharma2025quantum}
& \vspace{0.05cm}\centerline{\faDotCircleO}
& \vspace{0.05cm}\centerline{\faCircle}
& \vspace{0.05cm}\centerline{\faCircle} 
& \vspace{0.05cm}\centerline{\faDotCircleO}
& \vspace{0.05cm}\centerline{\faCircleO} \\
\hline
\centerline{2026} & 
\centering \cite{verma2026optical}, \cite{zhou2026chaotic}, \cite{lai2026fast}, \cite{peng2026injecting}, \cite{lai2026design}
& \vspace{0.05cm}\centerline{\faDotCircleO}
& \vspace{0.05cm}\centerline{\faDotCircleO}
& \vspace{0.05cm}\centerline{\faCircleO}
& \vspace{0.05cm}\centerline{\faCircle}
&  \vspace{0.05cm}\centerline{\faCircleO}\\
\hline
\end{tabular}
\vspace{0.2cm}
\footnotemark [1]{ Here,  \faCircle  $\ \rightarrow$ completely attained,  \faDotCircleO $\ \rightarrow$ partially attained, \faCircleO $\ \rightarrow$ not attained.}
\end{table}
\end{center}

In extensive video transmissions, essential security characteristics include ensuring confidentiality and data integrity, maintaining resilience and video quality, and preventing data loss. \hyperlink{table3}{Table 3} indicates that the current chaotic compression and encryption methods are unable to fully address all these security concerns. Among the numerous existing approaches, some lack sufficient time and space efficiency; others fail to uphold data integrity and confidentiality, while a few are inadequate in preserving the acceptable video quality. Consequently, as evidenced by \hyperlink{table3}{Table 3}, a novel chaotic methodology is essential to comprehensively tackle these challenges.

\subsection{Research Gap Analysis}
In any video communication system, the preservation of confidentiality, integrity, and video quality are key factors. Nevertheless, the discourse in this section demonstrates that the current methodologies do not adequately meet these criteria in a cohesive manner.  Articles \cite{akkasaligar2020medical, karmakar2021sparse} and \cite{shanmugam2025optimizing} present bio-inspired chaotic compression methodologies for video data processing. Nonetheless, these methodologies are insufficient for evaluating the frame rate in video processing. It diminishes the resilience and visual quality. The papers \cite{song2019efficient}, \cite{xu2020robust}, and \cite{el2022chaos} propose temporal composition and dictionary-based video compression methods. The inadequate temporal composition and absence of an efficient lexicon render these methods less successful. Effective communication through video data is essential for progress in metaverse and augmented reality-orientated applications \cite{chai2020efficient}, \cite{li2025error}. The primary concerns with these methodologies are elevated power consumption and computational difficulties. Conversely, the inadequate noise-handling mechanism and the complexity of temporal processing render the articles \cite{van2013encryption}, \cite{long2025improved} and \cite{chen2025security} significantly ineffective. The multi-level encryption and block-scrambling methods for protecting coloured video transmission are also vulnerable to various security risks and transmission faults \cite{hosny2023fast}, \cite{shifa2020mulvis}. The chaotic map-based block permutation methods for generating cryptic video techniques fail to ensure sufficient data security and privacy. Quantum computation is employed here to diminish time complexity; nevertheless, it fails to ensure sufficient data integrity \cite{shao2024multi}, \cite{verma2026optical}. The methods for providing diverse hardware-orientated encryption techniques to safeguard video processing and transmission over IoT networks are additionally hindered by issues related to video quality, data privacy, and data loss \cite{nazish2025cm}, \cite{ahuja2023iot}. Additionally, \hyperlink{table1}{Table 1}, \hyperlink{table2}{Table 2}, and \hyperlink{table3}{Table 3} demonstrate that the diverse existing chaotic methodologies do not meet the necessary criteria for video transmission over the internet and also do not preserve video quality during retrieval at the receiving end. Consequently, a sophisticated methodology is essential to meet all requisite criteria for the processing and transmission of substantial video content. 

\hypertarget{sec3}{}
\section{Proposed Technique}
The integration of chaos with concurrent compression and encryption offers a cohesive strategy for minimising data size and safeguarding content. The chaos approach incorporates randomisation into the compression process, guaranteeing that the compressed data is inaccessible without the appropriate key. This strategy mitigates the shortcomings of conventional sequential techniques, which frequently exhibit inefficiency and elevate computational demands. Initiatives to conserve space and bandwidth while enhancing security. Utilising a chaotic pseudo-random key stream and dynamically modifying the encoding technique based on the chaotic map, the video system will deliver effective compression and robust encryption. The chaotic entropy can be quantified using \hyperlink{eqn17}{Equation (17)} to evaluate the efficacy of any continuous chaotic encryption method. According to the convention, a strategy that yields a greater chaotic entropy value is deemed more efficient, and vice versa \cite{umar2024chaos, luca2004new}. If the chaotic entropy value of any security strategy is denoted as $\eta(\alpha)$ and $P(\alpha)$ represents the probability distribution, then Equation (17) can be expressed as follows.

\hypertarget{eqn17}{}
\begin{equation}\tag{17}
\eta(\alpha)= \int P(\alpha)\log P(\alpha) \ d\alpha
\end{equation}
According to Shannon's entropy theory, the entropy values for any chaotic map are significantly dependent on the sequences of the input data $(\alpha)$ in a one-dimensional chaotic map \cite{umar2024chaos}. Chaos-based technology is particularly suited for routine video applications, like streaming, surveillance, and real-time broadcasting, where both performance and security are paramount. Consequently, incorporating chaos into video compression and encryption signifies the initial phase towards scalable, secure, and efficient video management. The development of the proposed technique and its associated processes are illustrated in \hyperlink{Fig4}{Fig. 4}. 

\begin{figure}[htbp]\hypertarget{Fig4}{} 
\centerline{\includegraphics[scale =0.62]{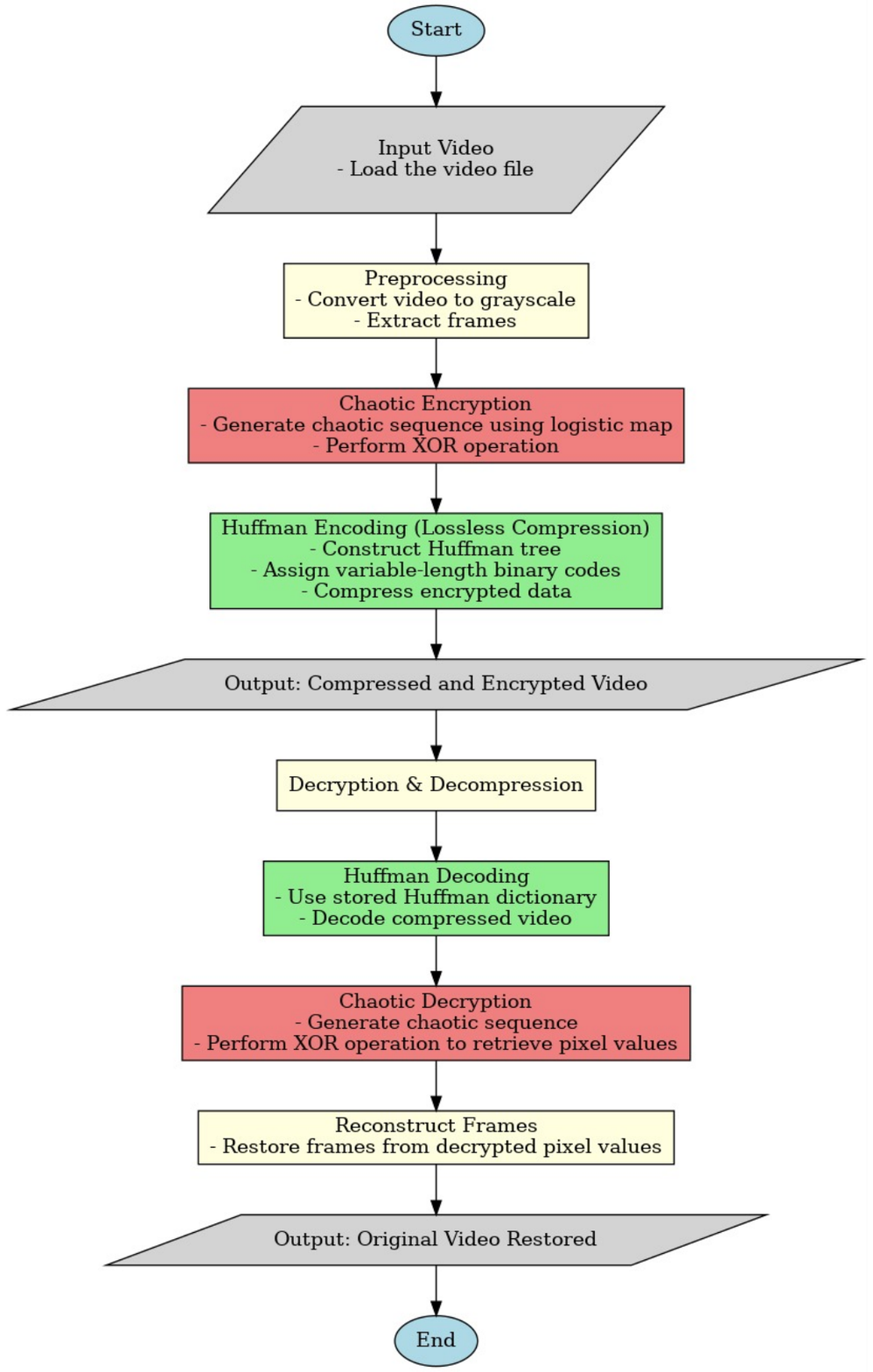}}
\caption {Workings of the proposed technique}
\end{figure}

\subsection{Encoding and Compression}
The encoding and compression of videos commence with the processing of the video and its conversion into a sequence of frames. The primary techniques following this involve encryption via a logistic chaotic map to generate randomness for improved security, the XOR operation to safeguard pixel values, and Huffman coding for lossless compression. Decryption and decompression entail reversing the processes of encryption and compression to ultimately restore the frames and get the original video. The chaotic S-box is generated in this procedure by employing \hyperlink{eqn6}{Euation (6)} and \hyperlink{eqn7}{Equation (7)} to replace the input data. Subsequently, adaptive Huffman compression is implemented on the replaced compressed data $(\delta)$. Simultaneously, the pseudorandom key generation process is utilised for the development of the secret key stream $(\rho)$; additionally, the resultant chaotic compressed and encrypted data $(\Delta)$ can be produced using \hyperlink{eqn18}{Equation (18)}.
\hypertarget{eqn18}{}
\begin{equation}\tag{18}
\Delta= \delta \oplus \rho
\end{equation}

\subsubsection{Chaotic Map Encryption}
Video processing begins with the loading of input video data and the extraction of essential information, including all frames and frame rates, to ensure synchronisation throughout encoding and decoding. Each frame is then converted to greyscale, reducing computational complexity while preserving critical information. This preliminary phase optimises subsequent activities, such as encoding/encryption and compression, hence improving performance while maintaining the video's usability. The encryption process utilises the chaotic behaviour of the logistic map to generate a pseudo-random sequence suitable for secure video encryption. This parameter selection ensures that the sequence is very sensitive to initial conditions, a hallmark of chaos that complicates replication and prediction. To align the intensity values of greyscale pixels, the generated sequence is scaled and quantised to the byte range (0, 255). The XOR operation is performed between the pixel values and the chaotic sequence values, with the output replacing each pixel value. This method preserves the safety and security of the pixel data, as the original data cannot be recovered without knowledge of the initial conditions and chaotic parameters, along with the generation of the chaotic sequence.

\subsubsection{Huffman Encoding for Lossless Compression}
Huffman encoding is a form of lossless compression that assigns variable-length binary codes to data symbols based on their frequency. More prevalent symbols will have abbreviated codes, while conversely, less frequent symbols will obtain longer codes. A Huffman tree is formed using this, which facilitates the preparation of the Huffman dictionary. This guarantees that the compression procedure is efficient and does not result in data loss. Let n represent the total quantity of pixels within the frame. If $\rho(\nu)$ represent the probability of the occurrence of a certain pixel value $\nu$. Let, $\Gamma$ represents the array of pixels; hence, $\rho(\nu)$ can be computed using the following \hyperlink{eqn19}{Equation (19)}. 

\hypertarget{eqn19}{}
\begin{equation}\tag{19}
\rho(\nu)=\frac{\text{count}(\nu)}{n}, \quad  \forall \nu  \in \text{unique}(\Gamma)
\end{equation}

The probability distribution $P(\nu)$ is utilised to construct the Huffman tree. Each distinct value is designated as a leaf node, with each node assigned a weight of $\nu$. Two nodes with the minimum weights are now amalgamated, referred to as 'parent nodes' $(PN)$, possessing a weight equivalent to the sum of the weights of the child nodes, which can be computed using the following \hyperlink{eqn20}{Equation (20)}. This procedure is reiterated until only one node persists to calculate the final weight of the parent node $(\omega (\eta))$. 

\hypertarget{eqn20}{}
\begin{equation}\tag{20}
\omega(\eta) = \rho(\nu_1) + \rho(\nu_2)+...+\rho(\nu_n)
\end{equation}

If binary code $\beta(\nu)$ is allocated to each pixel value, determined by the trajectory from the root of the Huffman tree to the leaf node. When $\rho(\nu)$ is substantial, a diminished value of $\beta(\nu)$ is allocated, and conversely, to guarantee effective compression. The overall proposed chaotic encryption and the compression technique is shown with the following \hyperlink{algo1}{Algorithm 1}.

\hypertarget{algo1}{}
\begin{algorithm}
\caption{Simultaneous Video Compression and Encryption}
\begin{algorithmic}[1]
\State \textbf{Input:} Video file $V$
\State \textbf{Output:} Compressed and Encrypted Video $V_{enc}$
\State \textbf{Initialize:} Chaotic map parameters $\nu$, $\lambda$
\State Read video file $V$, extract frames and frame rate
\State Create video writer for encrypted video

\For{each frame $F$ in $V$}
    \State Convert $F$ to grayscale
    \State Flatten $F$ into a 1D vector
    \State Initialize chaotic sequence $S$
    
    \For{each pixel $p$ in $F$}
        \State Update chaotic variable: $\nu =\lambda \times \nu \times(1-\nu)$
        \State Compute the chaotic sequence value: $S(p) = \lfloor \nu \times 10^8 \rfloor \mod 256$
    \EndFor
    
    \State Encrypt frame: $E = p \oplus S$
    
    \State Compute symbol probabilities
    \State Generate Huffman dictionary $D$
    \State Apply Huffman Encoding: $C = \text{HuffmanEncode}(E, D)$
    
    \State Save $C$ as encrypted frame in $V_{enc}$
\EndFor

\State \textbf{Return} $V_{enc}$

\end{algorithmic}
\end{algorithm}

\hyperlink{algo1}{Algorithm 1} demonstrates the functionality of the proposed simultaneous encryption and compression method. In \hyperlink{algo1}{Algorithm 1}, steps 1 and 2 delineate the input video file (V) and the resultant compressed and encrypted file ($V_{enc}$). The chaotic parameters ($\nu$ and $\lambda$) were initialised in step 3. Step 4 processes the incoming video file and extracts the frames $(F)$. Simultaneously, it computes the frame rate as well. A video writer has been developed in step 5 to compose the encrypted video. The generation of a compressed and encrypted video file, utilising each frame of the input video file $(V)$, has been accomplished between steps 6 and 19. Step 7 transforms each frame into a greyscale format, while step 8 compresses the greyscale frames $(F)$ into one-dimensional vectors. At this point, step 8 involves initialising the chaotic sequence $(S)$. Steps 9 to 13 modify the chaotic parameters based on each pixel value $(p)$ derived from the flattened frames $(F)$ and produce the final chaotic sequence.  Step 14 generates the final encrypted video file $(E)$ by executing the $XOR$ operation between the final chaotic sequence and the appropriate pixel values. Subsequently, Huffman compression is implemented on the disordered encrypted video file. In step 15, the probabilities of all unique symbols (pixel values) are computed, while step 16 builds the Huffman dictionary $(D)$ to store each unique symbol together with its related probabilities.  The disordered encrypted and Huffman-compressed file $(C)$ was produced by utilising the final encrypted video file $(E)$ and the Huffman dictionary $(D)$ as input parameters in step 17. The output video file ($V_{enc}$) has been produced in step 19 by saving the chaotic encrypted and Huffman compressed file $(C)$ into $V_{enc}$. Step 20 yields the output video file ($V_{enc}$) for transmission via any network.

\subsection{Decryption and Decompression}
Upon receipt of the complete video file at the receiving end, the entire chaotic encrypted and compressed video is regarded as the input for the original video retrieval methods. The quality of the returned video  will determine the evaluation of the merits of the proposed chaotic technique. Consequently, it is important to retrieve the original video from the received encrypted and compressed file. The original video is extracted from the compressed-encrypted version by reversing the approach. This stage involves first decompressing the incoming video file using Huffman decompression. Additionally, the decompression procedures based on chaotic maps are performed to recover the original video file.  The subsequent subsection delineates each method for recovering the original video.

\subsubsection{Huffman Decoding}
During the decoding procedure, the compressed codes are first segregated, followed by the creation of the associated dictionary for these codes. During this decompression phase, the construction of a Huffman tree is necessary to generate the dictionary matching to each compressed code. Consequently, akin to the data compression procedure, decompression uses the compressed code as input to reconstruct the original symbol using the Huffman tree. The decompression procedure necessitates the updating of the Huffman tree following the decoding of the original symbol, contingent upon data statistics. The chaotic, encrypted and compressed video has been used as input in this phase and decompressed using the same Huffman dictionary. The appropriate Huffman tree is navigated, producing the original encrypted values, hence ensuring lossless decompression and data integrity. The efficacy of the suggested technique may be assessed by calculating the average code length using \hyperlink{eqn16}{Equation (16)} and the total number of video pixels retrieved from the compressed file post-decompression \cite{usama2021efficient}. The total quantity of extracted video pixels may be determined using the following \hyperlink{eqn21}{Equation (21)} in conjunction with \hyperlink{eqn16}{Equation (16)}.

\hypertarget{eqn21}{}
\begin{equation}\tag{21}
Total \  Bits= T_{Comp} \times A_{CL}
\end{equation}

In \hyperlink{eqn21}{Equation (21)}, $T_{Comp}$ denotes the total quantity of compressed codes, while $A_{CL}$ signifies the average code length. According to the convention, if the total number of extracted video pixels from the decompressed file is equal to or nearly equal to the number of original input pixels, the employed technique is deemed efficient or vice-versa \cite{hermassi2010joint,usama2021efficient}. Consequently, in this phase, we have quantified the amount of extracted (encrypted) pixels and compared these figures with the original input pixels.

\subsubsection{Chaotic Map Decryption}
The decompressed video now comprises encrypted pixels, which must be decrypted to retrieve the original video. These encrypted pixels are taken as the inputs in this phase.  This can be accomplished with a chaotic sequence. The initial encrypted values acquired post-Huffman decoding are decrypted utilising the chaotic sequence employed during the encryption process \cite{wu2024novel, zhang2005image}. Using the previously reported chaotic values of $\lambda$ and $\nu_0$ from the \hyperlink{eqn7}{Equation (7)}, a chaotic sequence is constructed. If the decoder employs the identical reverse methodology as \hyperlink{eqn13}{Equation (13)} and \hyperlink{eqn14}{Equation (14)} to decode the encoded value, utilising \hyperlink{eqn22}{Equation (22)} and \hyperlink{eqn23}{Equation (23)}. 

\hypertarget{eqn22}{}
\begin{equation}\tag{22}
\oint'(\theta')= 
\left[ \frac{\theta'-\alpha_{i-1}}{\beta_{i-1}-\alpha_{i-1}} \right] \times \lambda 
\end{equation}

In \hyperlink{eqn22}{Equation (22)}, $\oint'(\theta')$ signifies the decryption function, utilising the chaotic parameters $\{ \theta', \lambda\}$ in conjunction with the pixel intervals $(p_i)$ defined by $\{\alpha_i, \beta_i\}$ for pixel numbers $(i)=\{1,2,3, \ldots, n\}$.

\hypertarget{eqn23}{}
\begin{equation}\tag{23} 
\theta' \in \Delta_i= \begin{cases}
    \left[ \frac{\theta'-\alpha_0}{\beta_0-\alpha_0}\right] \\ \\
    \left[ \frac{\theta'-\alpha_1}{\beta_1-\alpha_1}\right] \\
    \ \ \ \ \ \vdots \\ 
    \left[ \frac{\theta'-\alpha_n}{\beta_n-\alpha_n}\right]
\end{cases}
\end{equation}

\hyperlink{eqn23}{Equation (23)} delineates the intervals $(\Delta_i)$ pertinent to the decryption process of various pixels $(i=1,2,3, ...., n)$ utilising the decryption function $\oint'(\theta')$. This disorderly sequence will align with the one employed throughout the encryption procedure. The original pixel value is obtained by executing an XOR operation \cite{usama2021efficient} between the chaotic sequence value and the pixel's encrypted value. The procedures necessary to recover the original pixel values from the chaotic encrypted and compressed file at the receiving end are delineated in \hyperlink{algo2}{Algorithm (2)}.

\hypertarget{algo2}{}
\begin{algorithm}
\caption{Simultaneous Video Decompression and Decryption}
\begin{algorithmic}[1]
\State \textbf{Input:} Compressed and Encrypted Video $V_{enc}$
\State \textbf{Output:} Decrypted Video $V_{dec}$
\State \textbf{Initialize:} Chaotic map parameters $\nu$, $\lambda$
\State Read compressed video file $V_{enc}$
\State Create video writer for decrypted video

\For{each frame $C$ in $V_{enc}$}
    \State Decode using Huffman Decoding: $E = \text{HuffmanDecode}(C, D)$
    
    \State Initialize chaotic sequence $S$
    
    \For{each pixel $p$ in $E$}
        \State Update chaotic variable: $\nu = \lambda \times \nu \times (1 - \nu)$
        \State Compute chaotic sequence value: $S(p) = \lfloor \nu \times 10^8 \rfloor \mod 256$
    \EndFor
    
    \State Decrypt frame: $F = E \oplus S$
    \State Reshape $F$ into original 2D frame
    \State Save $F$ in $V_{dec}$
\EndFor
\State \textbf{Return} $V_{dec}$

\end{algorithmic}
\end{algorithm}

The simultaneous video decompression and decryption steps are performed using \hyperlink{algo2}{Algorithm 2}, using the received encrypted and compressed video file at the receiving end. In \hyperlink{algo2}{Algorithm 2}, steps 1 to 3 accept the input ($V_{enc}$), declare the output array for storing the retrieved original pixel values necessary for generating the output video, and initialise the chaotic parameters ($\nu$, $\lambda$) for executing the complete retrieval process. Step 4 processes the input encrypted and compressed video file ($V_{enc}$), while Step 5 establishes a video writer to generate the output video file containing the decompressed and decrypted pixel values. Steps 6 to 16 are employed to obtain the output video file ($V_{dec}$) by concatenating the retrieved original video frames ($F$).Step 7 involved the decompression and extraction of the original video ($E$) from the compressed video pixels using the Huffman decompression method, which requires two parameters: the compressed video pixel values ($D$) and the Huffman tree ($C$). Step 8 initialised the chaotic sequence ($S$). Steps 9 to 12 decrypt the encrypted pixels obtained from the decompression procedure. In Step 10, update the chaotic variable ($\nu$) with the chaotic parameter ($\lambda$) using the formula $\lambda \times \nu \times (\nu-1)$. In Step 11, calculate the chaotic pixel sequence $\left(S(p)\right)$ for each pixel value ($p$). Step 13 decodes the disordered pixel values to reconstruct the original frames using the formula $ F=E \oplus S$, while step 14 reformats all the recovered frames into two-dimensional structures. Ultimately, all the extracted, decompressed, and decrypted frames are stored in a video array $(V_{dec})$, and the proposed decompression and decryption procedure returns the array containing the extracted video frames $(V_{dec})$.

\subsubsection{Reconstruction of Frames}
Upon completion of the decryption procedure, the retrieved original pixel values are compiled in the corresponding matrices. The neighboring matrices of pixel values are subsequently transformed into the matching video frames. Subsequent to the reconstruction of video frames, these frames are re-organised and transformed into a raw video file. The output video file has been captured and transformed to match the format of the input video file. This reshaping method restores the original integrity of the movie and enables an exact reconstruction of the uncompressed raw footage.

\subsection{Potential Use Cases}
The suggested chaotic technique is applicable in various contexts, including online cloud-based video surveillance systems, where video data must be transmitted from one point to another \cite{huang2025efficient, shi2024heterogeneous}. The proposed method can guarantee the security and integrity of video data while transmission over any unsecured network, such as the internet. This strategy can be quite beneficial for organisations such as Vertex Plus, Hangzhou Hikvision Digital Technology, and Dahua Technology that manage distant video data. This approach is also relevant for drone-based remote video processing and surveillance mechanisms. Consequently, firms such as EOS GmbH, Shield AI, and ideaForge may likewise implement the proposed methodology.  Additionally, it can be utilised in telemedicine applications where distant video data serve as the major inputs for delivering specialised remote medical services. Consequently, organisations such as Teladoc, Amwell, Sesame Care, and other tele-medicine firms can implement these strategies.  This approach can be utilised for digital rights management, corporate and enterprise communication, law enforcement, covert operations, and other scenarios where compressed and encrypted video transmission is essential for maintaining data integrity and security \cite{pande2010joint, shanmugam2025optimizing}.

\subsection{Time Complexity Analysis}
The entire work has been conducted in the C++ programming language. To enhance processing efficiency, this research divides huge video files into fixed-size tiny video blocks and processes them sequentially in parallel. The research indicates that the time complexity of any chaotic compression and encryption method escalates with the rise in input video size \cite{ayubi2021new, tiwari2025compressed}. Nevertheless, owing to the fixed input video block sizes and concurrent execution, the time and space complexity will remain relatively stable.  Consequently, the implementation of both proposed encoding methods, namely encryption and compression, along with the decoding techniques, specifically decryption and the decryption process, may be performed in polynomial time. \hyperlink{algo1}{Algorithm 1} for encoding and \hyperlink{algo2}{Algorithm 2} for decoding demonstrate the utilisation of two for loops in both instances. Each time, the outer for loop accesses $n$ video frames, while the inner for loop accesses $m$ pixels, with it $n$ being significantly greater than $m$. Consequently, if each encoding and decoding operation associated with \hyperlink{algo1}{Algorithm 1} and \hyperlink{algo2}{Algorithm 2} requires $O(1)$ time complexity, then the overall encoding and decoding processes will each necessitate $O(n)$ time complexity to execute $n$-related operations. Additionally, both the encoding and decoding processes exhibit continuous time complexities, which are rather insignificant. Consequently, the cumulative time complexity for both encoding and decoding will be $O(n)$ the same. 

\hypertarget{sec4}{}
\section{Instances of Execution and Inspection}
This section encompasses two fundamental facets of this scientific endeavor. The initial section encompasses the Experiment Details, which provides information regarding the software configuration, hardware arrangement, and data preparation for evaluating the proposed chaotic compression and encryption, as well as assessing its performance across many dimensions. The second portion of this section encompasses assessment parameters that delineate various evolution matrices applicable for evaluating the efficacy of the proposed method in safeguarding data integrity and privacy.

\subsection{Experiment Details}
The implementation of the proposed chaotic methodology requires three fundamental components: specifications of the hardware configuration, software configuration, and data preparation details. Consequently, this subsection encompasses an in-depth examination of these three aspects.

\subsubsection{Software Setup}
This experiment was developed and carried out using the Ubuntu 22.04 LTS-based private cloud of UPES (Dehradun, India). The input and output data were stored in cloud-based storage, which was then utilised to evaluate the suggested chaotic approach's performance in several areas. Only the input and output video files, as well as their transfer from one place to another within the wireless communication network, have been stored in this storage area. The suggested method was constructed using the GCC-14 and the C++ programming language. To speed up the execution of the suggested strategy, a parallel execution mechanism was constructed using C++. \hyperlink{tableA1}{Table A1} in the \hyperlink{appendix}{Appendix} section lists the download sources for the different software files that are used.
\subsubsection{Hardware Setup}
Five sets of personal computer nodes, each with 2TB of secondary storage, were utilised for data transfer during the execution of the suggested chaotic technique. Moreover, each personal computer is equipped with 32GB DDR3 RAM and an Intel Core i8 processor. All data transfers between local computers and the remote cloud have been conducted via wired (local area network) and wireless (WiFi) connections. Ten terabytes of storage were utilised on the cloud platform for the retention of huge video data inputs and outputs. Owing to hardware resource limitations, the efficacy of the suggested chaotic methodology across several dimensions has been evaluated using a limited size of extensive video input data.
\subsubsection{Data Preparation}
The implementation of the proposed chaotic technique is constrained by certain hardware and software resource limits. Consequently, to address these restrictions, a maximum of $2TB$ of input video data was utilised in evaluating the performance of the proposed chaos technique. The 64 MB size is optimal for C++ processing. A distinct C++ application has been created to partition video files into 64 MB segments to enhance execution through an effective parallel processing approach.  A merger application is created to consolidate the extracted output video files at the receiving end. The .mp4 video files serve as input to evaluate the effectiveness of the suggested approach across various parameters. As a consequence, the unique videos are downloaded from various platforms and converted into the .mp4 format for use as inputs in this study.  The encoded video clips are subsequently amalgamated to render them appropriate for input via online video merger platforms. The origins of different sources of downloaded files and open-source software are tabulated  in \hyperlink{tableA1}{Table A1} in the \hyperlink{appendix}{Appendix} section.

\subsection{Assessment Parameters}
Three factors have been used to evaluate the suggested chaos video compression's performance: its ability to provide data integrity, data confidentiality, and time and space complexity reduction. Its ability to generate a better Peak Signal-to-Noise Ratio (PSNR) and percentage of information loss $(PIL)$ has demonstrated its effectiveness in providing higher data integrity. Its capacity to provide higher entropy values and avalanche effects has also been investigated in order to determine its potential to generate higher data privacy. The capacity to generate faster throughput has been used to test time efficiency, while the power to generate fewer bits per code has been used to test space efficiency. The following subsections provide a detailed definition of each performance metric.

\subsubsection{Peak Signal-to-Noise Ratio (PSNR)} This metric assesses the quality of regenerated or compressed video by comparing the original and processed signals in decibels $(dB)$, with higher values signifying enhanced fidelity \cite{wu2024novel, kaccar20224d}. In any 8-bit video, the significant values typically range from 30 $dB$ to 50 $dB$. The calculation is performed using \hyperlink{eqn24}{Equation (24)}, which incorporates the maximum pixel value $(\mu)$ and the mean squared error $(MSE)$, as determined by \hyperlink{eqn3}{Equation (3)}.
\hypertarget{eqn24}{}
\begin{equation}\tag{24}
PSNR= 10 \times \log_{10}{\left(\frac{\mu}{MSE}\right)} 
\end{equation}
Convention dictates that a higher $PSNR$ value from any video compression or processing method signifies enhanced quality of the resultant video. The proposed chaotic method operates on video pixels, with each pixel value ranging from 0 to 255, representable using 8 bits. Consequently, in any pixel-based video processing system, the $PSNR$ can fluctuate between 30 $dB$ and 50 $dB$.    
\subsubsection{Percentage of  Information Loss (PIL)}
During file transmission, information may be lost or distorted due to channel noise, interference from unauthorised third parties, or limitations of the transportation system \cite{bhattacharjee2024leveraging, bhattacharjee2016security}. Moreover, in certain instances of video data processing methodologies, a degree of information may be compromised throughout the retrieval process. Consequently, for these many reasons, the proportion of information loss in any chaos-based video compression algorithm is determined using \hyperlink{eqn25}{Equation (25)}.

\hypertarget{eqn25}{}
\begin{equation}\tag{25}
PIL= \frac{Input \ file \ size -Decompressed \ file \ size}{Input \ file \ size} \times 100
\end{equation}

According to the conversion, any chaotic compression method is deemed efficient if it results in a smaller percentage of information loss or the opposite.

\subsubsection{Entropy}
Information theory states that the uncertainty associated with a random variable can be determined using entropy. Entropy is essentially the cryptic that is injected into the video data in any chaotic-based encryption method \cite{bhattacharjee2024leveraging, umar2024chaos}. Consequently, a greater entropy number indicates that there is less illegal information in any unsecured video data. According to Shannon's theorem, \hyperlink{eqn26}{Equation 26} may be used to determine the entropy $(\eta(\alpha))$ of any given source ($\alpha$).

\hypertarget{eqn26}{}
\begin{equation}\tag{26}
\eta(\alpha)=\sum_{i=1}^{2n-1}{\rho(\alpha_i) \times \log_{2}{\frac{1}{\rho(\alpha_i)}}}
\end{equation}

In \hyperlink{eqn26}{Equation 26}, $\rho(\alpha_i)$ denotes the probability of the symbol $\alpha_i$. According to Shannon's theory, the optimal entropy number for an encrypted movie should be 8. Per the consensus, a cryptographic method with an entropy value of 8 is deemed the most robust for safeguarding against various security threats compared to alternative ways.

\subsubsection{Avalanche Effect}
The alteration of one or several bits during chaotic encryption, decryption, or transmission can result in significant variations in the produced video. Consequently, the output video may be entirely altered. The encrypted video data is often indecipherable due to the alteration of one or more bits. Consequently, in any chaotic encryption, the avalanche effect $(AE)$ demonstrates how a minor alteration in the input video during encryption might influence the resultant ciphertext. It additionally computes the quantity of bits influenced in the ciphertext if a single bit is altered \cite{bhattacharjee2023simultaneous}. The avalanche effect can be expressed by  \hyperlink{eqn27}{Equation 27} as follows.

\hypertarget{eqn27}{}
\begin{equation}\tag{27}
AE=\frac{Total \ number \ of \ flipped \  bits \ in \ an \ output \ cryptic \ video \ file }{Total \ number \ of \ bits \ in \ an \ input \ video \ file}
\end{equation}
According to \hyperlink{eqn27}{Equation 27}, $AE$ is utilised to assess the robustness of any encryption method in safeguarding against transmission faults or security threats. If a cryptic algorithm yields a higher $AE$, it indicates that the employed encryption technology is very efficient or robust against specific transmission mistakes and security assaults, or conversely.

\subsubsection{Throughput}

In a data processing system, each specific task must be executed within a designated timeframe. The duration needed to do various tasks with different execution mechanisms differs based on the characteristics of the execution mechanism and the diversity of the input file type and size. Throughput is defined as the quantity of work accomplished within a specified time interval, as noted in references \cite{usama2017chaos,bhattacharjee2015lossless}. In addition, \hyperlink{eqn28}{Equation 28} is used to model the throughput. 
\hypertarget{eqn28}{}
\begin{equation}\tag{28}
Throughput= \frac{Output \ video \ file \ size}{Total \ execution \ time}
\end{equation}

According to \hyperlink{eqn28}{Equation 28}, a chaotic video data processing system that provides increased throughput is deemed time-efficient, and vice versa.

\subsubsection{Bits per code}
In any chaos-based video data compression method, bits per code (BPC) denotes the quantity of output bits necessary to represent a pixel value in a compressed output file \cite{bhattacharjee2024leveraging, usama2021efficient}. BPC can be articulated by the \hyperlink{eun29}{Equation 29}.

\hypertarget{eqn29}{}
\begin{equation}\tag{29}
BPC=\left(\frac{Output \ compressed \ video \ file \ size}{Input \ file \ size}\right) \times 8
\end{equation}
Here in \hyperlink{eqn29}{Equation 29}, 8 represents the pixel length in bits. The efficacy of any chaos-based video compression is deemed efficient for file size reduction if it provides a reduced bits per code ($BPC$).

\hypertarget{sec5}{}
\section{Result Analysis}
This section examines the performance of the proposed chaotic video encryption and compression technology in three aspects: data integrity, information loss, and data privacy. The efficacy of the proposed technique in these areas is further contrasted with the performance of other existing similar approaches to demonstrate its dominance over them.

Data integrity is a crucial component of any data transfer system. It may be impeded by numerous transmission failures, security breaches, constraints of the transmission technology, and other factors. Typically, such events result in the corruption of one or more bits, leading to failures in the data retrieval process and resulting in partial or whole data loss. This phenomenon is basically known as information loss. Consequently, data integrity may be compromised due to these factors. According to the literature, the data integrity may be computed by determining the $PSNR$ values and assessing the percentage of information loss using \hyperlink{eqn24}{Equation 24} and \hyperlink{eqn25}{Equation 25}. The efficacy of the suggested strategy, along with other similar methods, in generating $PSNR$ values and percentage of information loss $(PIL)$ is illustrated in \hyperlink{Fig5}{Fig. 5} and \hyperlink{table4}{Table 4}.

\begin{figure}[htbp]\hypertarget{Fig5}{} 
\centerline{\includegraphics[scale =0.35]{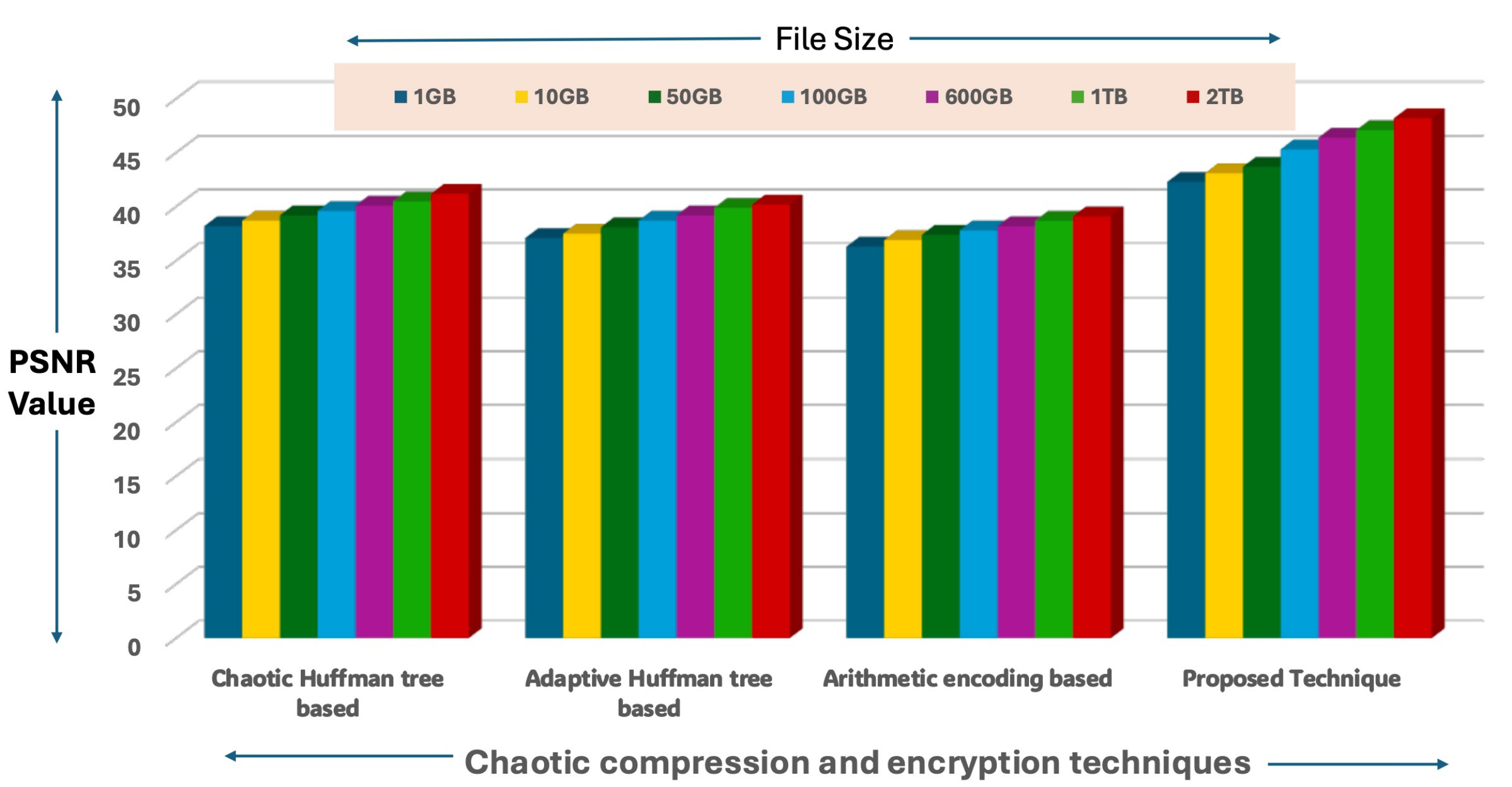}}
\caption {$PSNR$, offered by different chaotic approaches}
\end{figure}

\hyperlink{Fig5}{Fig. 5}  illustrates that among the limited number of established chaotic methods already employed, the proposed chaotic encryption and compression technique yields the highest PSNR values. Consequently, according to the definition of $PSNR$, the suggested method, it provides superior video output quality during the retrieval of video frames at the receiving end. Moreover, the literature \cite{gomes2025end} indicates that superior output video quality can be preserved if any security method mitigates transmission failures and minimizes data loss during transmission, processing, or retrieval. Consequently, we assert that the proposed strategy is effective in minimizing transmission errors and data loss during processing and transmission, relative to other comparable methods, thereby validating its ability to improve data integrity.  

According to the definition, any data processing and transmission system that ensures a reduced percentage of information loss is guaranteed to preserve data quality \cite{bhattacharjee2024leveraging}. 
Consequently, the proposed chaotic strategy, along with other current methods, ensures data quality and integrity, which is further examined in \hyperlink{table4}{Table 4}.      

\begin{table}[htbp]
\centering
\small
\renewcommand{\arraystretch}{1.5}
\setlength{\tabcolsep}{3pt}
\hypertarget{table4}{}
\caption{Percentage of Information Loss, offered by different techniques}
\resizebox{\textwidth}{!}{%
\begin{tabular}{|p{4.5cm}|c|c|c|c|c|c|c|p{3.2cm}|}
\midrule
\multirow{3}{*}{} & \multicolumn{7}{c|}{} & \multirow{5}{*}{\centerline{$\Bigg\uparrow$}}\\

\multirow{0.5}{*}{\centerline{\textbf{Chaotic compression \&}}} & \multicolumn{7}{c|}{\textbf{File Size}} & \multirow{6}{*}{\centerline{\textbf{Percentage of }}}\\

\multirow{0.5}{*}{\centerline{\textbf{encryption techniques}}} & \multicolumn{7}{c|}{\textbf{}} & \multirow{6}{*}{\centerline{\textbf{Information Loss}}}\\

\cline{2-8}
& \textbf{1 GB} & \textbf{10 GB} & \textbf{50 GB} & \textbf{100 GB} & \textbf{600 GB} & \textbf{1 TB} & \textbf{2 TB} & \\
\cline{1-8}
Chaotic Huffman tree based & 0.0082 & 0.0081 & 0.0079 & 0.0075 & 0.0069 & 0.0067 & 0.0065 & \multirow{5}{*}{\centerline{$\Bigg\downarrow$}}\\
\cline{1-8}
Adaptive Huffman tree based & 0.0091 & 0.0089 & 0.0083 & 0.0079 & 0.0076 & 0.0071 & 0.0069 & \\
\cline{1-8}
Arithmetic encoding based & 0.0098 & 0.0093 & 0.0089 & 0.0086 & 0.0083 & 0.0079 & 0.0075 &  \\
\cline{1-8}
\textbf{Proposed Technique} & \textbf{0.0040} & \textbf{0.0037} & \textbf{0.0031} & \textbf{0.0028} & \textbf{0.0021} & \textbf{0.0018} & \textbf{0.0009} &  \\
\hline
\end{tabular}%
}
\end{table}

\hyperlink{table4}{Table 4} analyses the efficacy of various chaotic methods in minimizing information loss, including the proposed strategy. These analyses of performance in lowering the proportion of information loss have been conducted for various video input data set sizes.   \hyperlink{table4}{Table 4} demonstrates that the suggested chaotic encryption and compression methodology exhibits a reduced percentage of information loss compared to the other relevant chaotic methods stated. This fact indicates that the proposed strategy is the most effective in minimizing information loss across all scenarios. Consequently, as illustrated in \hyperlink{Fig5}{Fig. 5} and \hyperlink{table4}{Table 4}, the proposed technique demonstrates superior efficiency in enhancing data integrity compared to other related methods, particularly in reducing information loss and improving $PSNR$ values consistently.      

The quality of the retrieved video further verified by comparing the input video frame and the retrieved video frame at \hyperlink{Fig6}{Fig. 6}.  If there are no perceptual discrepancies between the original input video frame and the recovered video frame, the employed video processing method is deemed error-prone and effective in minimizing data loss. Consequently, the similarities and the differences between the original and retrieved video frames is illustrated in \hyperlink{Fig6}{Fig. 6}.

\begin{figure}[htbp]\hypertarget{Fig6}{} 
\centerline{\includegraphics[scale =0.44]{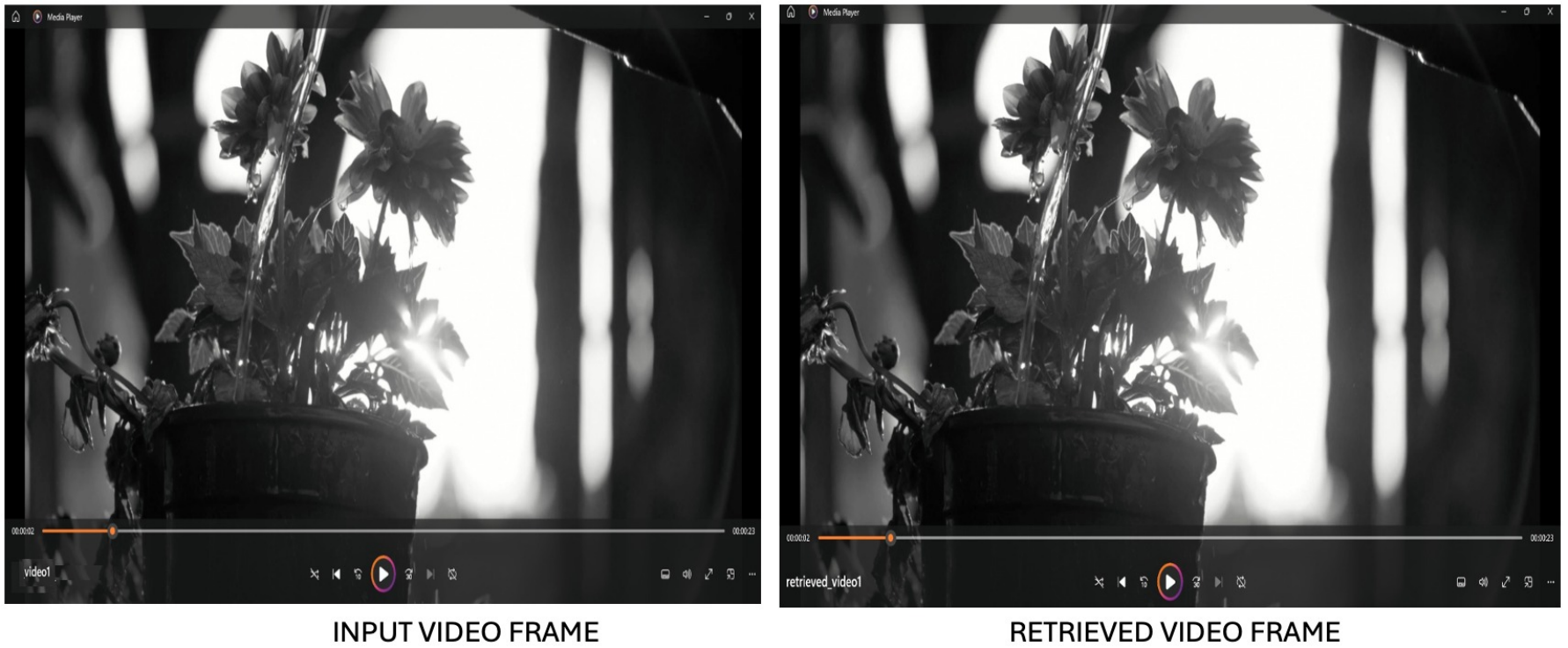}}
\caption {$PSNR$, offered by different chaotic approaches}
\end{figure}

\hyperlink{Fig6}{Fig. 6} illustrates that there is no discernible difference between the original input video frame and the retrieved output video frame. Consequently, based on \hyperlink{Fig6}{Fig. 6}, we assert that the proposed technique provides enhanced data integrity. Therefore, from the  \hyperlink{Fig5}{Fig. 5}, \hyperlink{Fig6}{Fig. 6} and  \hyperlink{table4}{Table 4} justify the first objective of this research work.

The integrity of the data may potentially be compromised due to temporal and spatial overheads. Increased time and space complications may lead to heightened channel congestion and overhead, perhaps resulting in the loss of transmitted video data. The space complexity of any video processing method can be diminished by effective data compression, while time efficiency can be improved via intelligent data processing and the implementation of parallel processing techniques. 
Consequently, the suggested chaotic method effectively integrated data encryption and compression to provide enhanced temporal and spatial efficiency. Its facilitation of parallel processing for video input data enhanced its time efficiency. Literature indicates that a video data processing system is considered time-efficient if it provides larger throughputs \cite{usama2017chaos}. Likewise, if any data compression method yields fewer bits per code, it is deemed space-efficient \cite{usama2021efficient}.  This section computes the throughputs and bits per code generated by the proposed chaotic encryption and compression technique, as well as other similar chaotic methods, utilising \hyperlink{eqn28}{Equation 28} and \hyperlink{eqn29}{Equation 29}.  The resultant throughputs and bits per code produced by these chaotic methods are subsequently illustrated in \hyperlink{table5}{Table 5} and \hyperlink{fig7}{Fig. 7}.

\begin{table}[htbp]
\centering
\small
\renewcommand{\arraystretch}{1.2}
\setlength{\tabcolsep}{4pt}
\hypertarget{table5}{}
\caption{Throughputs, offered by different chaotic techniques}
\resizebox{\textwidth}{!}{%
\begin{tabular}{|p{4.7cm}|p{4.8cm}|p{3.2cm}|}
\midrule
\centering\textbf{Chaotic compression and encryption techniques} &
\centering\textbf{Throughputs (MB/Sec) of Chaotic Encryption and Compression} &
\centering\textbf{Throughputs (MB/Sec) of Retrieval} \tabularnewline
\hline
Chaotic Huffman tree based & 2.71 & 2.84 \\
\hline
Adaptive Huffman tree based & 3.13 & 3.21 \\
\hline
Arithmetic encoding based & 3.41 & 3.57 \\
\hline
\textbf{Proposed Technique} & \textbf{4.61} & \textbf{4.75} \\
\hline
\end{tabular}%
}
\end{table}

\hyperlink{table5}{Table 5} presents the throughputs associated with the encryption and compression of input video data, in addition to the throughput for data retrieval. Thus, it demonstrates that the proposed chatty methodology consistently yields superior outputs compared to other relevant methodologies. Consequently, the suggested chaotic method is more time-efficient than the other alternatives.   

\begin{figure}[htbp]\hypertarget{Fig7}{} 
\centerline{\includegraphics[scale =0.40]{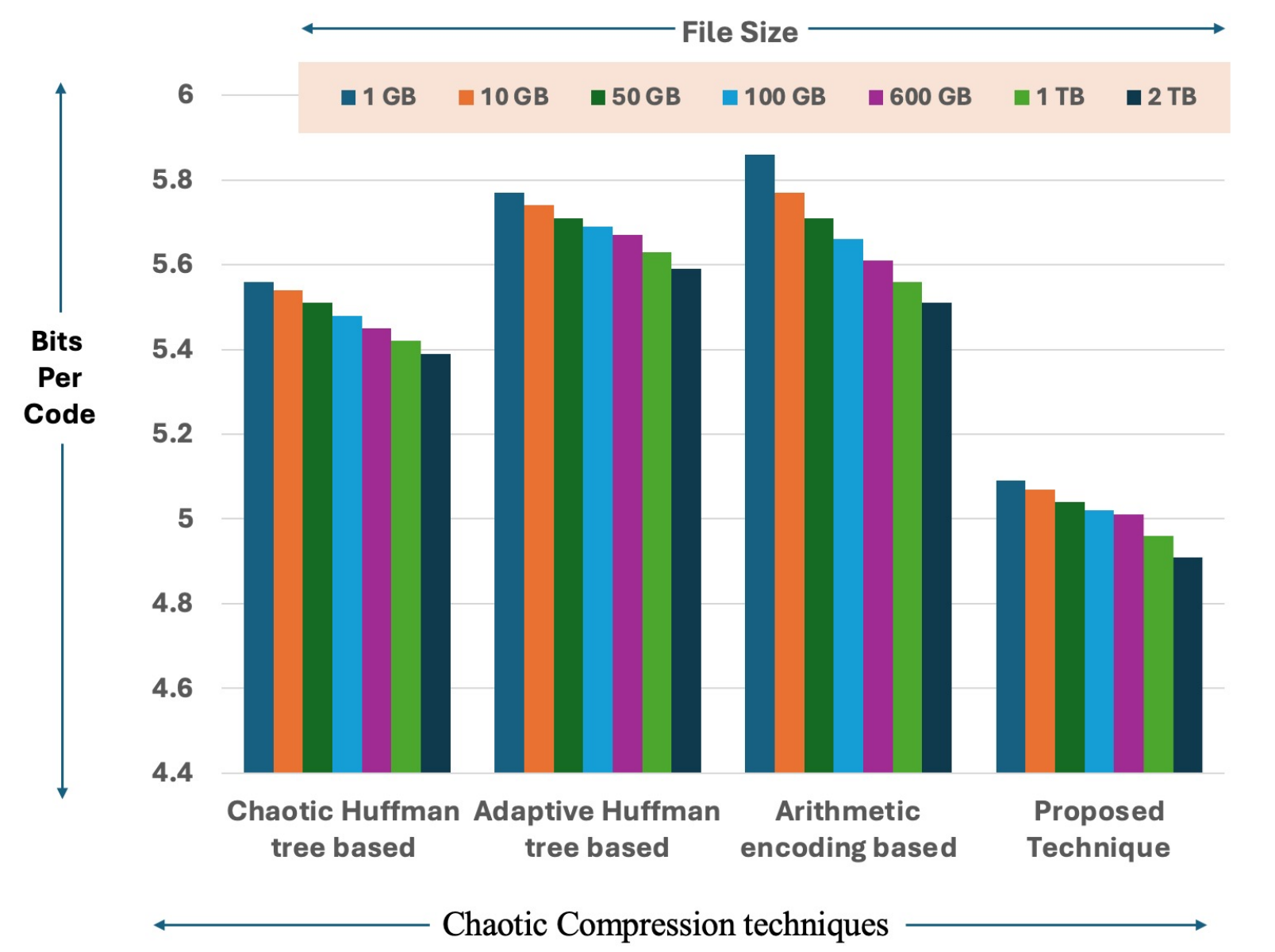}}
\caption {Bits Per Code, offered by different chaotic compression approaches}
\end{figure}

In \hyperlink{fig7}{Fig. 7}, the suggested chaotic method demonstrates a reduced number of bits per code compared to other related approaches in nearly all instances. Consequently, the proposed chaotic strategy is effective in minimizing space overheads compared to previous comparable methods. Consequently, \hyperlink{table5}{Table 5} and \hyperlink{fig7}{Fig. 7} demonstrate the superior time and space efficiency of the suggested method compared to other current chaotic approaches, thereby validating the second purpose of this research.

Data secrecy is a critical component of any data transmission system. This section assesses the data confidentiality provided by the proposed chaotic encryption and compression methodology, as well as other related methods, regarding their ability to deliver elevated entropy levels and significant avalanche effects. Utilising \hyperlink{eqn26}{Equation 26} and \hyperlink{eqn27}{Equation 27}, the entropy values and avalanche effects are computed and presented in \hyperlink{table6}{Table 6}.

\begin{table}[htbp]
\centering
\small
\renewcommand{\arraystretch}{1.2}
\setlength{\tabcolsep}{4pt}
\hypertarget{table6}{}
\caption{Avalanche effect and entropy values offered by different chaotic techniques}
\resizebox{\textwidth}{!}{%
\begin{tabular}{|p{7.5cm}|c|c|}
\hline
\centering\textbf{Chaotic compression and encryption techniques} &
\centering\textbf{Avalanche Effect (in \%)} &
\centering\textbf{Entropy Values} \tabularnewline
\hline
Chaotic Huffman tree based  & 63.2 & 7.63 \\
\hline
Adaptive Huffman tree based & 64.3 & 7.53 \\
\hline
Arithmetic encoding based   & 62.2 & 7.67 \\
\hline
\textbf{Proposed Technique} & \textbf{76.2} & \textbf{7.77} \\
\hline
\end{tabular}%
}
\end{table}

\hyperlink{table6}{Table 6} demonstrates that the suggested technology provides superior entropy values and avalanche effects compared to other current chaotic-based encryption and compression methods. Thus, \hyperlink{eqn26}{Equation 26} and \hyperlink{eqn27}{Equation 27} indicate that a security technique demonstrating elevated entropy values and pronounced avalanche effects is deemed efficient in providing enhanced data confidentiality. Consequently, \hyperlink{table6}{Table 6} substantiates that the proposed technique is effective in providing enhanced data secrecy, thereby fulfilling the third research aim.   

\hypertarget{sec6}{}
\section{Conclusion}
The most crucial component of any data transfer system is guaranteeing data integrity and confidentiality. Data confidentiality and integrity are frequently jeopardised by a number of security risks, transmission system constraints, channel noise, channel congestion, and other elements.  These issues get more worse as the amount of data being sent grows. Transferring enormous volumes of video data while preserving its integrity and privacy and reducing the percentage of data loss is especially challenging.  Nevertheless, the current literature does not address this topic in a comprehensive manner. The closely related encryption and compression techniques are examined and reviewed in this paper. We also present new results from chaos theory, compression methods, and cryptography. Since there is no efficient integrated solution, we propose a technique based on the Huffman encoding and chaotic substitution-box algorithms. The performance of the suggested method to provide greater data confidentiality has been evaluated in this study using its ability to provide higher avalanche effects ($76.2\%$) and entropy values (7.77), and it has been compared with the other related methods currently in use to demonstrate its superiority over them. Its ability to provide a greater percentage of information $PSNR$ ($48.2 dB$ as max) and a lower percentage of information loss ($0.0009 \%$ as min) in comparison to other current approaches has been examined for its ability to guarantee data integrity. In addition, it exhibits temporal and spatial efficiency by achieving elevated throughputs for the generation of chaotic cryptic video (4.61 $MB/Sec$) and retrieval (4.75 $MB/Sec$) through parallel processing techniques, while employing a minimal bit rate (4.91 bits as min) to mitigate information loss and enhance data integrity and confidentiality. Consequently, the proposed approach is appropriate for comprehensively addressing all these concerns, thereby satisfying the objectives of this research.

\hypertarget{sec7}{}
\section{Future Work}
The findings section validates the benefits of the proposed chaotic encryption and compression strategy to enhance data integrity, data privacy and lower the percentage of information loss when compared to other comparable techniques already in use. Future research will concentrate on refining the suggested approach, and any alternative workable chaos theory-based simultaneous data encryption and compression solution would be greatly appreciated. Besides compression, forthcoming research and developments of the suggested work can augment cryptographic capabilities. The literature examines the merits and drawbacks of current chaotic encryption and compression methods to safeguard the transport of substantial video data. The suggested method provides a feasible and consistent approach to address diverse security challenges associated with the processing and transmission of substantial video data across unreliable platforms. The generation of pseudo-random keys by the amalgamation of several bits necessitates further investigation to enhance both security and computational efficiency. Consequently, the space complexity can be further enhanced by employing a more efficient data compression method.

\clearpage                 
\pagestyle{plain}
\bibliographystyle{unsrt} 
\bibliography{Manuscript_ref}

\newpage
\hypertarget{appendix}{}
\section*{Appendix}
\renewcommand{\arraystretch}{1.5}
This section outlines the programme specifications and their respective download sources. It also encompasses the download sources for various input files.  

\hypertarget{tableA1}{}
\begin{table}[h!]
\renewcommand\thetable{A1}
\caption{Download Sources of input files and software}
\begin{tabular}{|c|c|} 
\midrule 
\textbf{Input Files}  &  \textbf{Download Sources}\\ 
\hline
Video File &  https://www.pexels.com/search/videos/repository/ \\ 
 & https://coverr.co/ \\
Online .mp4 converter &	https://cloudconvert.com/mp4-converter\\	
GCC-14 & https://packages.debian.org/trixie/gcc-14-x86-64-linux-gnu\\
Video Merger  & https://www.adobe.com/express/feature/video/merge\\
& https://clideo.com/merge-video \\
Ubantu 22.04 LTS & https://releases.ubuntu.com/22.04/\\
\bottomrule
\end{tabular}
\end{table}

\end{document}